\documentclass[acmsmall]{acmart}

\AtBeginDocument{%
  }

\setcopyright{acmlicensed}
\acmJournal{PACMHCI}
\acmYear{2025} \acmVolume{9} \acmNumber{7} \acmArticle{CSCW382} \acmMonth{11}\acmDOI{10.1145/3757563}

\usepackage{xcolor}
\usepackage{subcaption}
\usepackage{float}
\usepackage{pgfplots}
\pgfplotsset{compat=1.17}

\newcommand{\designname}{\textit{Burst }}
\newcommand{\appname}{Burst}

\begin{document}

\title{Burst: Collaborative Curation in Connected Social Media Communities}

\author{Yutong Zhang}
\email{yutongz7@stanford.edu}
\affiliation{%
  \institution{Stanford University}
  \city{Stanford}
  \state{California}
  \country{United States}
}

\author{Taeuk Kang}
\email{taeuk@stanford.edu}
\affiliation{%
  \institution{Stanford University}
  \city{Stanford}
  \state{California}
  \country{United States}
}
\author{Sydney Yeh}
\email{sydxyeh@stanford.edu}
\affiliation{%
  \institution{Stanford University}
  \city{Stanford}
  \state{California}
  \country{United States}
}
\author{Anavi Baddepudi}
\email{Anavib@stanford.edu}
\affiliation{%
  \institution{Stanford University}
  \city{Stanford}
  \state{California}
  \country{United States}
}
\author{Lindsay Popowski}
\email{popowski@stanford.edu}
\affiliation{%
  \institution{Stanford University}
  \city{Stanford}
  \state{California}
  \country{United States}
}
\author{Tiziano Piccardi}
\email{piccardi@stanford.edu}
\affiliation{%
  \institution{Stanford University}
  \city{Stanford}
  \state{California}
  \country{United States}
}
\author{Michael S. Bernstein}
\email{msb@cs.stanford.edu}
\affiliation{%
  \institution{Stanford University}
  \city{Stanford}
  \state{California}
  \country{United States}
}

\renewcommand{\shortauthors}{Yutong Zhang et al.}

\begin{abstract}
  Positive social interactions can occur in groups of many shapes and sizes, spanning from small and private to large and open. However, social media tends to binarize our experiences into either isolated small groups or into large public squares. In this paper, we introduce \textit{Burst}, a social media design that allows users to share and curate content between many spaces of varied size and composition. Users initially post content to small trusted groups, who can then ``burst'' that content, routing it to the groups that would be the best audience. We instantiate this approach into a mobile phone application, and demonstrate through a ten-day field study (N=36) that Burst enabled a participatory curation culture. With this work, we aim to articulate potential new design directions for social media sharing.

\end{abstract}



\begin{teaserfigure}
  \centering
    \includegraphics[width=\textwidth]{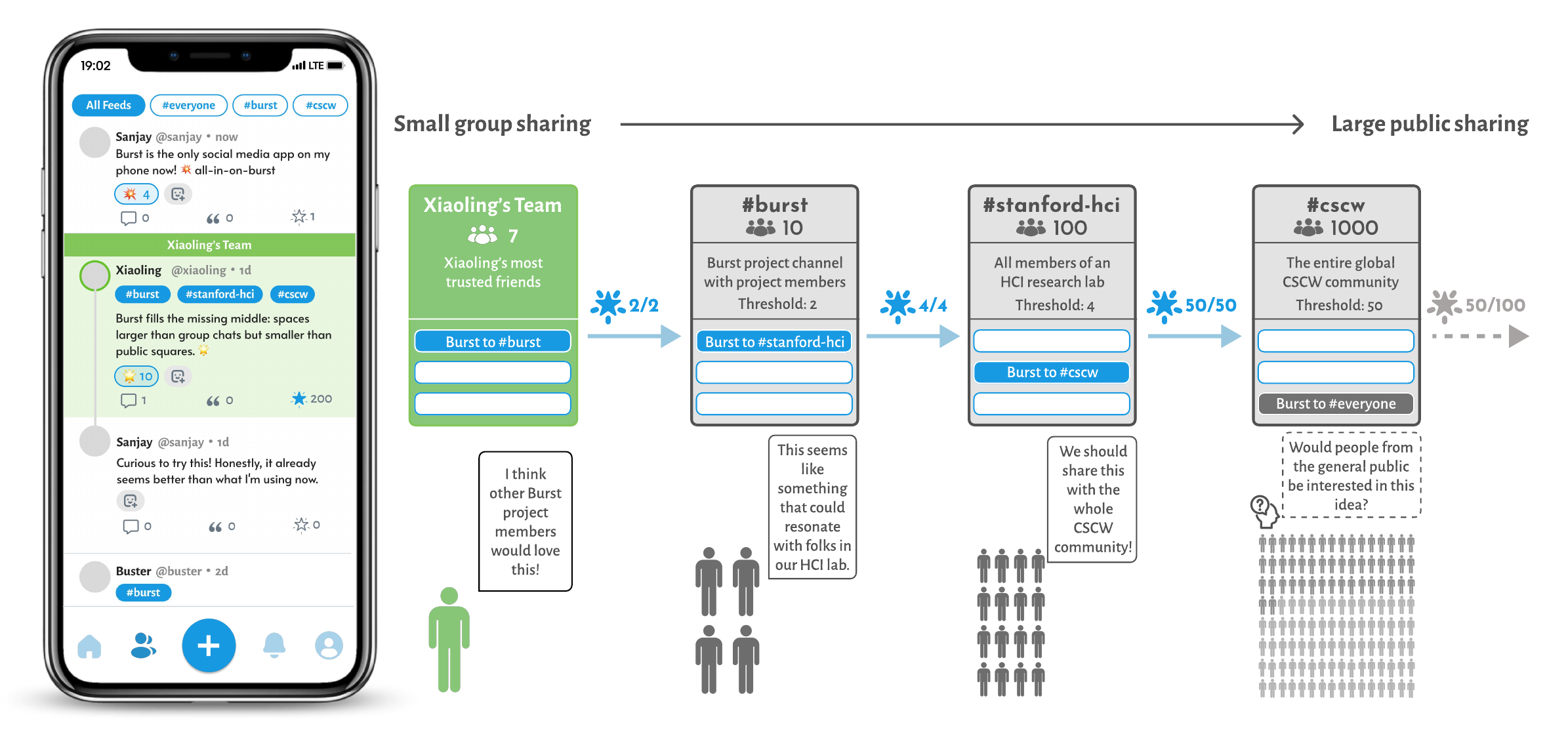}
  \caption{Burst is a social media design that centers post curation as an intentional and socially distributed activity. Posts begin visible only to a small group of trusted friends. Users curate posts that they think other groups would be interested in by \textit{bursting} posts from group to group, cascading the content to groups of different sizes and focuses as appropriate. A post only bursts into a group when enough users individually burst it enough to put it over the threshold set for that group. }
  \label{fig: main-burst-figure}
\end{teaserfigure}

\received{October 2024}
\received[revised]{April 2025}
\received[accepted]{August 2025}

\maketitle

\section{Introduction} \label{sec: introduction}

Social media was originally envisioned as a space for connection: a plurality of spheres where people could reach out, share diverse perspectives, and engage in meaningful dialogue~\cite{rheingold2000virtual,kapoor2018advances,boyd2007social}. Today, however, platforms have fractured into two dominant modes: large public squares (e.g., Twitter/X, TikTok) and small private silos (e.g., WhatsApp, Discord)~\cite{zhang2024form, rajendra2021illustrated}--- but neither of which live up to that original vision. 
While public squares are intended to support idea exchange and exposure to diverse perspectives~\cite{le2020town,habermas1991structural}, in practice, they are increasingly dominated by a small number of highly visible voices~\cite{soliman2024mitigating, lowry2006impact, kaligotla2016impact, gearhart2015something}, leaving the majority of users hesitant to contribute~\cite{das2013self}. Overcrowded with performative content and polarized exchanges, these environments often lead users to fear judgment, misinterpretation, or reputational risk~\cite{barasch2014broadcasting, marwick2011tweet, le2020town, beam2018context}. As a result, many people shifted their engagement to smaller, private spaces, such as group chats and closed communities, that can offer greater psychological safety, intimacy, and more predictable audience boundaries~\cite{rajendra2021illustrated}. However, while these private environments foster safe and intimate interactions, they also reduce exposure to diverse perspectives~\cite{soliman2024mitigating}, reinforce homophily~\cite{ferreyra2022community}, and risk becoming disconnected from broader public discourse. This fragmentation leaves us with diverse but isolated conversations in private groups, while public spaces are dominated by a vocal minority, resulting in polarized discourse and stalled cross-community engagement~\cite{viegas1999chat, grudin1994groupware, alexander1977pattern}.

This hollowing out of social media is an ecosystem-level challenge and a classic prisoner's dilemma issue in social computing without a neat solution~\cite{grudin1988cscw}. At the same time, our social interactions are much more than just small private conversations or fully public debate---so why should our social media reflect only those two poles? Would it be possible to create designs that more robustly fill out the \emph{missing middle} of spaces that are larger than small group chats but smaller than global public squares? Unfortunately, the people who would need to participate to enable such spaces are the ones least incentivized to post~\cite{grudin1988cscw}: the benefits of a more robust middle are diffuse and indirectly distributed, while the risks, including social judgment and reputational harm, are immediate and personally borne.

Consequently, content that might have value beyond a private group is either never created or never shared outside of it, and a broader ecosystem fails to self-sustain as a healthy active environment.

Platforms have added design features and algorithmic ranking features to mitigate this issue, but many have faced limited success because their features are still ultimately designed around either small or public audiences. For example, Instagram's Close Friends feature can hybridize a trusted private space within a public environment~\cite{elyukin2021behind}. Similarly, platforms such as Google+'s Circles~\cite{kairam2012talking} and Facebook’s audience group selection feature~\cite{gonccalves2011groupster} have been proposed to enable users to customize the visibility of their posts, tailoring content to different audiences. These audience control strategies intentionally and permanently block off the conversation from the larger public, like having a coffee shop miles away from the town center. Other strategies go the other direction, keeping groups isolated but encouraging crossposting~\cite{butler2012cross,boyd2010tweet}. Although these strategies enable sharing with a broader audience, they are often seen as counter-normative, accompanied by an implicit apology (e.g., ``Apologies for cross-posting''). Algorithms are a third possible strategy, for example trusting that TikTok's For You algorithm will direct the content to those who are interested. However, here the platform remains fundamentally public, and there is no stable notion of who the audience is for each post, so identity work remains difficult in the face of potential context collapse. Ultimately, the underlying design architectures of social media~\cite{zhang2024form} are at issue: we must rethink our design options.

To create a robust ecosystem of mid-sized communities that act as connective layers between the small group to the public square, in this paper we explore the concept of foregrounding \emph{content curation}---the act of intentionally redirecting content to new audiences---as the central design affordance of a social media platform. Those who consume the most content online are also those who are the most interested in curating interesting content for others to see~\cite{bernstein2010enhancing}: can we make this curation a main behavior? This approach not only facilitates content dissemination but also strengthens the connective tissue between subcommunities by enabling users to shape how and where content circulates.
Our approach contrasts with current prevailing platform paradigms in which curation is handled passively and invisibly by algorithmic systems, removing users from the social process of sharing content and from any notion of who can see the content.

To explore this alternative model, we introduce \emph{Burst}, a system that centers post curation as an intentional and socially distributed activity. Users can burst posts---curate via voting to share them---to another group, allowing both content and users to flow across groups with bursts serving as their bridge.
The design goal is for Burst to enable users to collaboratively share, frame, and route content toward the most appropriate audience spaces, helping bridge fragmented silos.
For example, a user might post a thought on CSCW research to their closest colleagues, with the intent that those colleagues will vet the thought on their behalf and only share it further if they think it is high quality. Their colleagues do, sharing it with the broader lab, where the post gathers positive feedback. From there, assuming the original user has not disabled broader sharing of the post, two lab members may jointly burst the post into the larger CSCW community. Later, a member of the lab might burst another piece of content shared from the larger CSCW community down into the lab group for smaller, private discussion. 

Burst otherwise operates as a threaded~\cite{zhang2024form} platform such as Twitter/X, Mastodon, Bluesky, and Threads---featuring a main feed of text and image posts, but adopting spaces as its central metaphor rather than network or commons~\cite{zhang2024form} more similarly to Facebook Groups, Reddit, and Discord---specifically organizing the platform into channels such as \#cscw, \#hci, and a platform-wide \#everyone channel. Initially, all new posts are shared only with a private channel of the user's most trusted friends. 
By bursting, those trusted friends vote to route content into other channels that the poster has enabled, with more votes required to burst into larger channels. 
The result is a collaborative curation~\cite{he2023cura} platform. 

Burst addresses several key challenges. First, it provides a safe space to experiment and fail, reducing the fear of negative public exposure. Second, it supports intermediate levels of success by allowing content to find medium-sized groups and visibility levels, rather than forcing binary outcomes of virality or privacy. Third, it normalizes the practice of managing audience boundaries and content direction as an expected, low-barrier part of participation. By centering the labor of curation and designing affordances that scaffold content direction, Burst aims to transform social media into a multi-layered, connection-centric ecosystem. 

To evaluate how Burst shapes online social behavior, we conducted a ten-day field study with 36 participants, who were asked to use the platform as they would any other social media. We analyzed data from participation logs, pre- and post-study surveys, and interviews to understand the social dynamics that arose on Burst. Participants actively engaged with the burst mechanism, sharing content across multiple channels and constructing connections between different audience groups. Participants reported feeling more confident and less hesitant to post, as they trusted their peers to help filter, moderate, and direct their content to the appropriate audience.

In summary, this paper contributes (1) a new social media design centered around sharing content between many small and large spaces; (2) the Burst mobile phone application, which implements this design in the context of a traditional text and image social media platform; and (3) a field experiment to understand the social dynamics arising from Burst.

\section{Related Work} \label{sec: related_work}
\subsection{Large public audience vs. Small groups}

Since their early development, social media platforms have evolved from and been influenced by the public nature of blogs and forums, coming to be perceived as digital public spaces for self-expression~\cite{shirky2008here,boyd2008youth,papacharissi2010networked}. The core design of platforms like Myspace, Facebook, Twitter, Reddit, and more recently, TikTok, Instagram, and YouTube maintain this assumption and incentivize people to use the platform as a ``public square''~\cite{boyd2007social}. The open design, which allows users to reach a wide audience, has positioned these platforms as digital spaces supporting public discourse~\cite{boyd2010social}, giving voice to everyone~\cite{papacharissi2010private} and enabling interactions with people who were previously out of reach~\cite{marwick2011see}. This design has accelerated information sharing and created the potential for viral content through social curation mechanisms, such as re-shares, which help disseminate information across broader networks~\cite{bakshy2012role,jenkins2013spreadable}. This wider reach of information has proved to be a design pattern with significant positive advantages, offering increased exposure to a diversity of perspectives~\cite{bakshy2015exposure}, supporting civic movements~\cite{tufekci2017twitter,castells2015networks}, and creating opportunities for public accountability~\cite{bennett2012logic}.

However, a social media design oriented toward a large public audience can have downsides. Expressing opinions publicly can pose privacy risks by revealing private information and increasing the risk of harassment~\cite{thomas2022s}. At the same time, posting content for a large audience often alters the type of content produced, encouraging self-presentation and the reframing of content to appeal to as many people as possible~\cite{marder2016strength,barasch2014broadcasting,taylor2022authentic, kaligotla2016impact}. One behavior that can be exacerbated by public audience design is the tendency to self-censor when opinions does not align with the mainstream ideas of the community~\cite{noelle1974spiral}. This tendency can result in a few loud voices dominating the conversation and a larger proportion of users refraining from active participation, thus potentially increasing the number of passive users. Additionally, a large public audience encourages rapid turnover of information, potentially leading to less engaged discussions~\cite{lorenz2019accelerating,tausczik2019impact}. Furthermore, virality does not apply only to valuable content: low-quality content also has an opportunity to spread widely~\cite{del2016spreading,vosoughi2018spread}, potentially causing harmful societal outcomes. Given the large volume of posts typical of large public groups, platforms must rely on algorithmic curation to select content. These algorithms can incorporate biases that can increase exposure to more extreme content creating risks for polarization and radicalization. The potential wide reach of poor-quality or harmful posts forces platforms to enforce moderation strategies that raise complex questions about trade-offs in freedom of expression and governance~\cite{gillespie2018custodians, mahar2018squadbox}.

Recent years have thus observed a rebalancing from large public audience platforms back to more private, intimate spaces for communication~\cite{boccia2021below,dzimianski2019isg15,srivastav2017going,velten2017managing,hwang2021people}. These communities include WhatsApp, Discord, iMessage, and Snapchat, which offer an experience more akin to a living room than to a public square. These platforms allow users to tailor their communication to specific groups, often resulting in more personal and engaged interactions~\cite{swart2019sharing}. 
Small groups offer several advantages, such as higher engagement, greater trust, a shift away from self-promotion, and a tendency to encourage thoughtful content sharing~\cite{rajendra2021illustrated}. 
Private conversations can lead to more focused, in-depth discussions where the focus on personal relations with the people in the groups makes it easier to estimate the reliability of information and reduce misunderstandings by providing the full context~\cite{tausczik2019impact}. Additionally, small private groups offer a degree of self-moderation by design, helping to contain low-quality or harmful content within the group boundaries. However, this design also has drawbacks. It can create silos, confining information within specific groups and reducing the diversity of perspectives. Additionally, limiting the reach can reduce opportunities for broader societal discussions.

\subsection{Connecting between Small Groups and Large Public Audiences on Social Media}

Effective integration of small groups and large public audiences faces several key challenges, including algorithm opacity, community identity dilution, and limitations in personalization and privacy. These challenges highlight gaps in existing mechanisms, which, while aiming to bridge narrow and broad communication modes, often fall short in providing a seamless user experience. 

On social media designed as large public spaces, algorithmic curation bridges the gap between audiences by allowing content to reach broad visibility. However, one major issue is the opacity of these algorithms: platforms like Facebook, Instagram, and TikTok use algorithms to tailor content based on users' engagement histories. Known as algorithmic ranking, this mechanism curates information to align with individual interests, creating a personalized experience. 
However, these algorithms are designed with different objectives that prioritize engagement metrics. The lack of transparency and customization of these algorithms intensifies the challenge, as users have little control over what content is prioritized or hidden from their feeds~\cite{barasch2014broadcasting,eg2023scoping}.

Another challenge is the dilution of community identity, which becomes particularly pronounced with crossposting. Crossposting allows content intended for one audience to be shared across multiple platforms or communities, reshaping community boundaries and broadening content reach \cite{butler2012cross,farahbakhsh2016characterization}. However, this practice can dilute the unique identity of individual groups, especially when external content clashes with the norms or values of a particular audience. For instance, on Reddit, where sub-communities (i.e., subreddits) often have distinct cultures, crossposting content from one subreddit to another can create friction and misalignment, as content that resonates in one community might be out of place or unwelcome in another~\cite{sawicki2023reddit}. 

Lastly, limitations in personalization and privacy settings impact the effectiveness of tools designed for more intimate sharing. Instagram's Close Friends, for example, enable users to share personal, authentic content with select audiences~\cite{xiao2020finsta, elyukin2021behind}. 
While these features allow users to curate their sharing, they remain isolated from the platform’s larger audience, thus restricting the reach and impact of personal narratives. This separation between large audience and small groups content restricts these tools from contributing meaningfully to larger social conversations, limiting their role in fostering public discourse. 

\subsection{Alternative Approaches to Social Media Design}

There are many known design patterns and strategies for how to organize the basic interactions of a social network~\cite{rajendra2021illustrated,zhang2024form}. Platforms like Mastodon, which operate on federated networks, provide an alternative social media design by allowing users to choose specific communities while still interacting across a broader network. This federation model, which rose to prominence as a result of Twitter moderation concerns, enables users to have more autonomy over the privacy and governance of their content. However, federated platforms still face challenges related to a complex mental model (``the fediverse'') and inability to effect any collective governance of antisocial behavior~\cite{robertson2019mastodon}.

Researchers have likewise proposed alternative designs for social media. We can reconsider identity mechanisms: for example, meronymous communication enables users to selectively disclose parts of their identity, aiming to foster more inclusive and diverse participation in public online spaces~\cite{soliman2024mitigating}. Likewise, synthesized social signals expand the set of ``honest signals'' that cannot be easily faked on a profile~\cite{im2020synthesized}. A second theme reconsiders who is in control, and when: for example, design approaches might prioritize explicit consent rather than consent-assumed designs when sharing, tagging, or otherwise engaging with participants publicly~\cite{im2021yes}. Third, social media might articulate alternative ranking algorithms that, for example, integrate societal objectives and not just engagement~\cite{jia2024embedding}. Finally, at the governance level, alternative technical and decision-making mechanisms might help social media achieve more effective governance and legitimacy~\cite{jhaver2023decentralizing,schneider2021modular,zhang2020policykit}. We take inspiration from each of these efforts. Burst is an attempt to expand the design space of sharing mechanisms, which are foundational to social media. With this intervention, we hope to lay the foundations for a more robust social media ecosystem and the ability to integrate solutions such as those above. 

\section{Burst}

Our design goal is to create a platform that populates an intermediate layer between private groups and public spheres---spaces where content can emerge from trusted groups and circulate to broader, yet bounded, audiences.
Grudin's Paradox explains why this sort of middle-layer sharing might fail to emerge: in collaborative systems, the individuals who benefit from contributions are typically not the ones who bear the costs~\cite{grudin1988cscw}. Applied to social media, this asymmetry discourages broader sharing. Users face immediate, personal risks such as judgment or reputational harm from sharing outside of trusted groups, while the benefits of contributing such as a more robust ecosystem are diffuse and delayed. As a result, much content either remains siloed or never gets posted.

Burst aims to mitigate this asymmetry by designing the platform itself around content curation. 
Rather than requiring individuals to make high-stakes sharing decisions alone, it allows members of a private space to collectively route the content to appropriate broader channels. This mechanism lowers the burden of contribution, since people can initially post to a private space with knowledge that their trusted friends will only curate it to a broader audience if appropriate. Likewise, it enables platform members to engage in behaviors beyond just liking or replying, with curation a meso-tier type of contribution. Finally, it allows content to move across communities without defaulting to full public exposure.

In this section, we begin by outlining the design goals behind \designname, explain how the overall design approach functions, and then explain how this design is realized in the Burst App for mobile phones.

\subsection{The Burst social media design}

Our approach aims to connect multiple small group communities with each other and with the larger public, to facilitate this middle social space.
Small-group social media platforms are effective in allowing users to share content within specific, targeted groups~\cite{hwang2021people}, but are often isolated from each other. Conversely, platforms designed for a broader audience enhance the rapid diffusion of information across an entire network~\cite{beam2018context}, but can lead to problems such as context collapse~\cite{marwick2011tweet}, which may discourage users from sharing content. Our goal is to build out the roadways between small private spaces and larger public spaces. Many feel more comfortable sharing with small, trusted groups~\cite{vitak2012impact}, and we further hypothesize that they trust those groups to curate which of their content ought to be seen by others. 

Our approach requires two basics: \textit{channels}, which are bounded groups of people, and \textit{thresholds} for each channel, which represent the minimum support needed for content to be shared with a channel. For example, we might have a channel representing a small group of students in a research lab, which only requires a threshold of one vote from a member to share new content into the channel. Or, we might have a large channel, \#hci, for global conversations of human-computer interaction, which has a higher threshold to mirror its larger size. 

\textit{Bursting}, then, is nominating content toward the threshold for a channel. A piece of content bursts into the channel when it reaches the threshold, and is withheld from the channel until then. The threshold is determined by a function based on the size of the channel: smaller channels require fewer bursts for content to successfully propagate, whereas larger channels require more. In other words, bursting links channels to one another and to the broader public by passing content around. A piece of content might start out in a small channel for students in a group, then burst out to all HCI students at the university, then burst out to the larger \#hci channel, then get bursted back down into a series of smaller channels as members of \#hci share it with their own local groups as well.

Unlike most large social media platforms, this approach does not allow content to be posted directly to a global feed, or even directly posted to specific channels: all posts begin visible to only a small group of trusted friends, who are the only ones empowered to curate that post into channels. In our implementation, the global feed (called \#everyone) has a higher threshold than any other channel, making it more difficult to reach everyone. Instead, all content is posted to smaller, trusted groups initially, which have lower burst thresholds.
While permitting users to directly post to channels---private or public---may offer greater flexibility and autonomy, we deliberately chose to restrict this functionality in order to lean into a simple sharing mental model: you post, your friends curate.
This decision also has the side effect of protecting groups from receiving a large number of low-effort posts; we depend on those users' friends to support the posters and curate only the appropriate content into larger channels.
By requiring content to first circulate in smaller, trusted groups, we aim to support safer initial expression, enable audience-sensitive curation, and reduce the spread of low-relevance or low-quality content to broader audiences.
We encourage the creation of many small group channels, which helps users control their imagined audience~\cite{marwick2011tweet}, ensuring they feel enough sense of safety and control over their sharing environment and have enough diversity to support specific self-presentation. Unlike common network-based sharing, where likes or shares broadly propagate content through a user’s social graph, bursting enables users to specify and constrain the audience for each post, allowing content to flow intentionally across community spaces rather than diffusely across networks.

In principle, this approach affords many benefits of social media curation. By requiring burst thresholds, the approach helps individual channels maintain norms and reduce anti-social behavior compared to town-square style social media~\cite{he2023cura}. By increasing thresholds for larger channels, this same can hold even with larger groups. By allowing users to suggest channels or block channels, individuals can control distribution if desired. Finally, to help ensure that channels themselves do not become havens for anti-social behavior, channels can be nested into a federal governance structure~\cite{schneider2021modular} to enforce rules and the curation thresholds could become algorithmic in nature and enforce bridging threshold algorithms~\cite{tornberg2023simulating} if needed.

Unlike traditional social media mechanisms---such as retweets, cross-posts, or algorithmic recommendations---Burst is explicitly designed to scaffold intentional, peer-mediated content flow across social boundaries. Retweeting and cross-posting often function as unilateral acts, where the original sharer or the re-poster takes full responsibility for directing content to a broader audience. These actions may appear abrupt, lack audience specificity, and are frequently accompanied by social risks or disclaimers. Algorithmic recommendations, on the other hand, outsource curation to opaque systems that offer limited user control and often prioritize engagement over contextual relevance. In contrast, Burst makes curation a socially distributed process, allowing content to circulate gradually and deliberately through collective endorsement.

\subsection{The Burst Application}
\begin{figure}[tb]
  \centering
  \includegraphics[width=\textwidth]{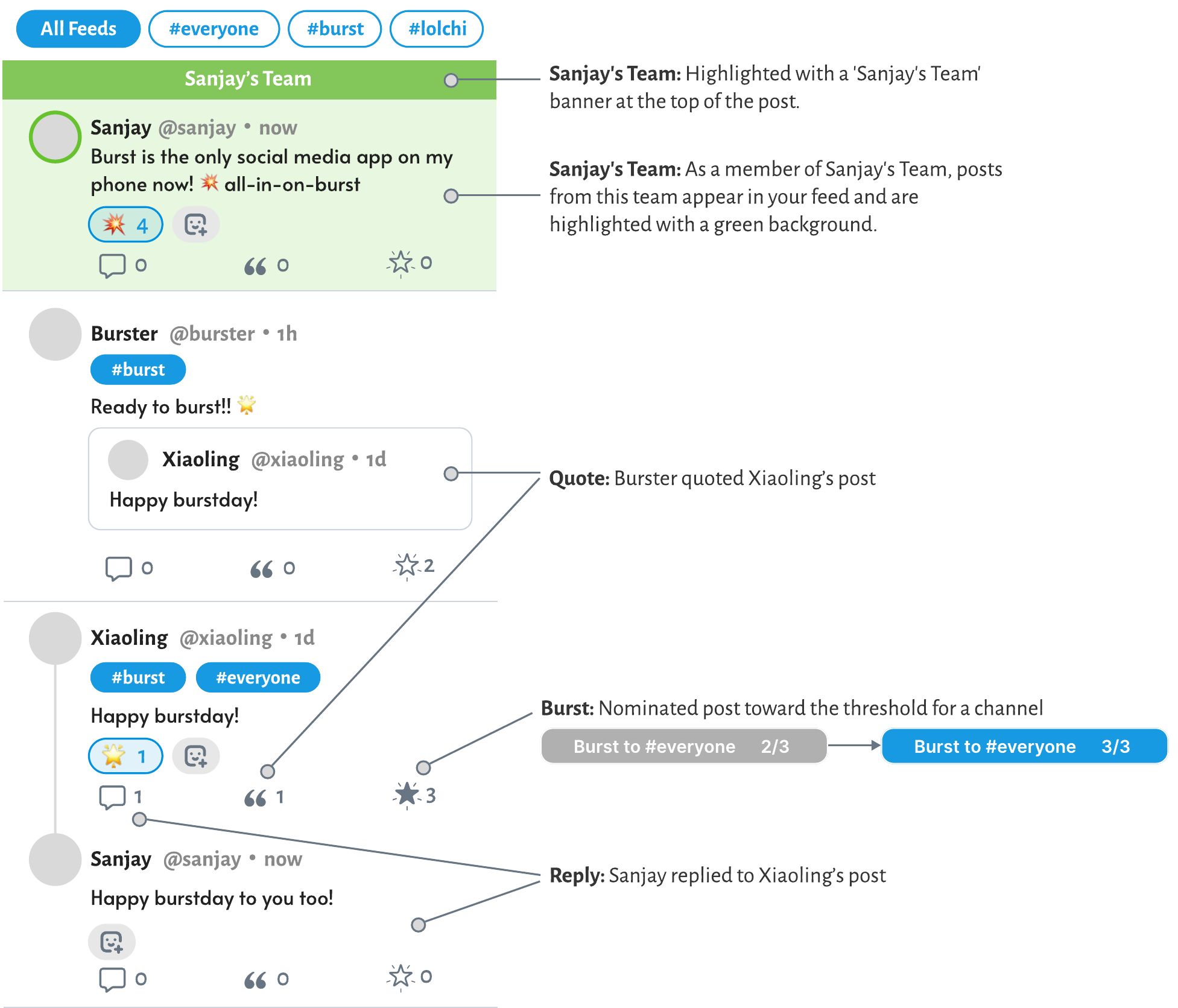}
  \caption{All feed views in Burst display posts in reverse chronological order, with the newest posts appearing at the top. The “All Feeds” page aggregates content from all channels and teams the user belongs to. For example, in this user’s sample “All Feeds” view, the feed includes a post from Sanjay, visible because the user is a member of Sanjay’s team, along with posts from the \#burst and \#everyone channels, which the user has joined. Users can reply to, quote, or burst any post. Additionally, users may filter the feed by selecting a specific channel from the top menu (e.g., \#everyone), which updates the view to display only posts that have successfully burst into that channel.}
  \label{fig:burst-feed}
\end{figure}

In principle, bursting could be integrated into many different kinds of social computing applications. For example, our initial prototype implemented bursting within Slack, where users could share content across channels using a custom emoji reaction. Our main system, however, develops the concept through a threaded-space~\cite{zhang2024form} social media application, which aligns most closely with the general usage patterns of mainstream social media platforms.\footnote{Threaded-space designs allow us to construct a feed featuring posts from channels that a user is following, where each post can receive comments and reshares, similar to hybridizing Reddit-style subreddits onto Twitter/X, Mastodon, Bluesky, or Threads.} Our application, named \appname{} (of course---we're not \textit{that} creative), demonstrates the underlying concepts. 
We chose to develop Burst as a standalone mobile app to fully isolate and evaluate the core mechanics of the bursting mechanism. This design decision enabled us to implement key features, such as “Your Team,” visual indicators of burst thresholds, and mechanisms for suggesting or blocking channels, that would not be feasible within existing platforms. By replicating familiar features from mainstream social media, we reduced potential confounds and made it easier for participants to understand and engage with the burst mechanism without confusion or context collapse from unrelated platform dynamics.
In this section, we explain the app design, as well as the community structure created for our field study. 
\appname{} is realized as a mobile application with a React Native frontend and a Node.js backend, which we deployed on iOS using TestFlight and on Android using Android Studio for distribution via Google Play.

\begin{figure}[htbp]
  \centering
  \begin{subfigure}[t]{0.28\textwidth}
    \centering
    \includegraphics[width=\textwidth]{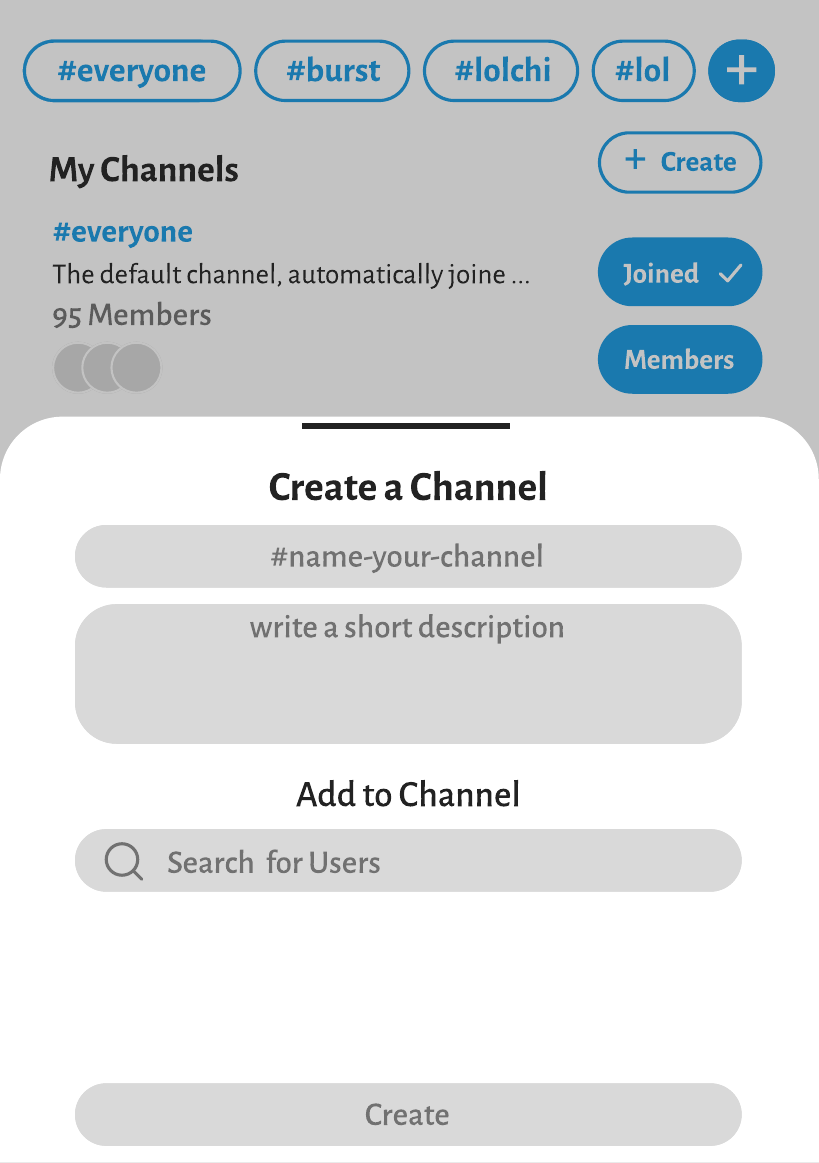}
    \caption{\textbf{Channels:} Users can join or create channels, with posts burst into channels they belong to appearing on their main feed. Users can also burst posts from their feed into any channel they have joined.}
    \label{subfig: community}
  \end{subfigure}
  \hspace{0.010\textwidth}
  \begin{subfigure}[t]{0.28\textwidth}
    \centering
    \includegraphics[width=\textwidth]{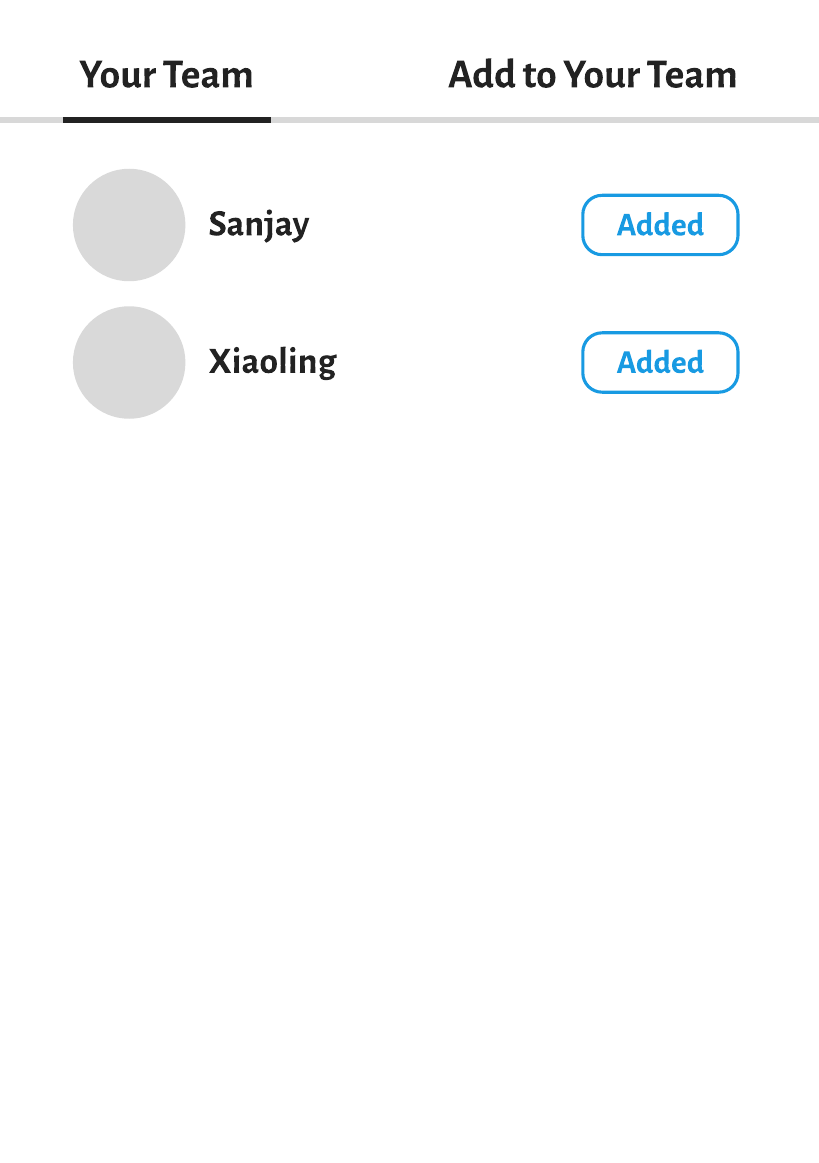}
    \caption{\textbf{Your Team:} Users can also invite others to join their team, which is an initial audience for post review. These team members serve as a private space for sharing content before broader dissemination.}
    \label{subfig: your_team}
  \end{subfigure}
  \hspace{0.010\textwidth}
  \begin{subfigure}[t]{0.28\textwidth}
    \centering
    \includegraphics[width=\textwidth]{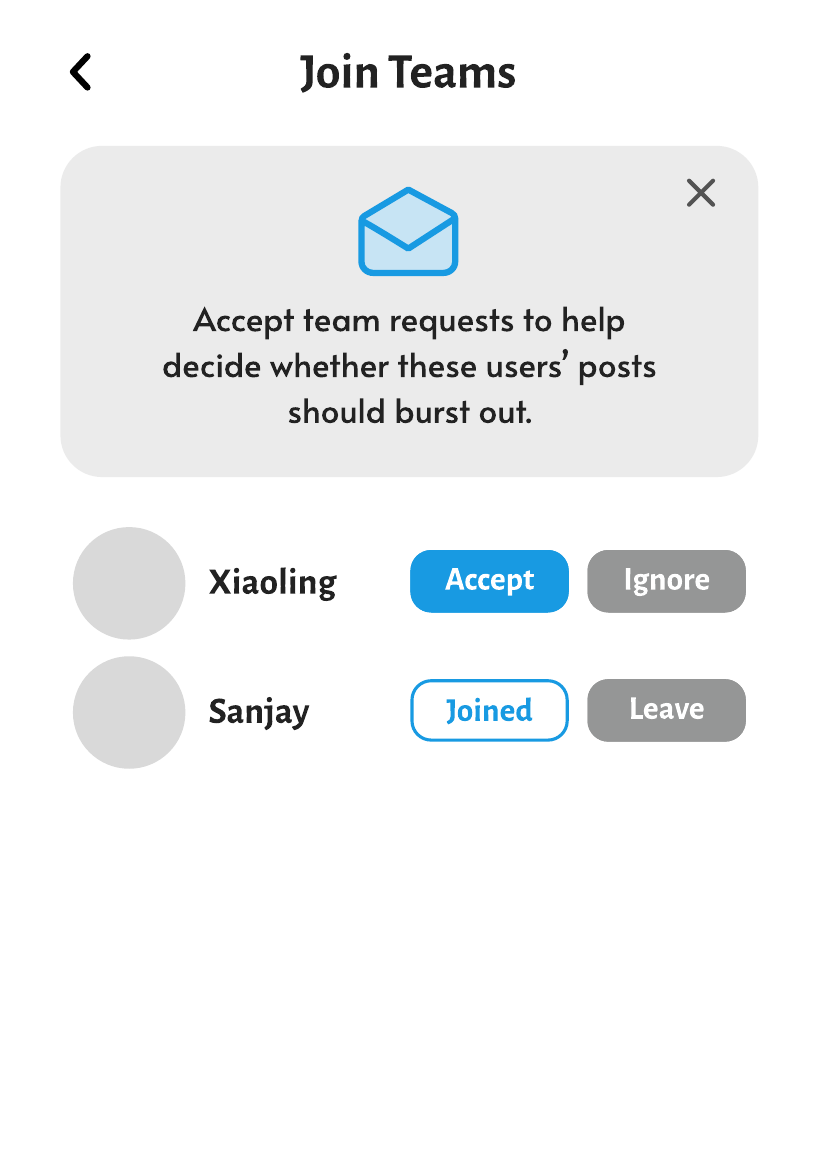}
    \caption{\textbf{Joined Team:} Users can accept others' invitations to join their team, becoming part of the initial audience and receiving notifications whenever they post.}
    \label{subfig: join_team}
  \end{subfigure}
  \begin{subfigure}[t]{0.28\textwidth}
    \centering
    \includegraphics[width=\textwidth]{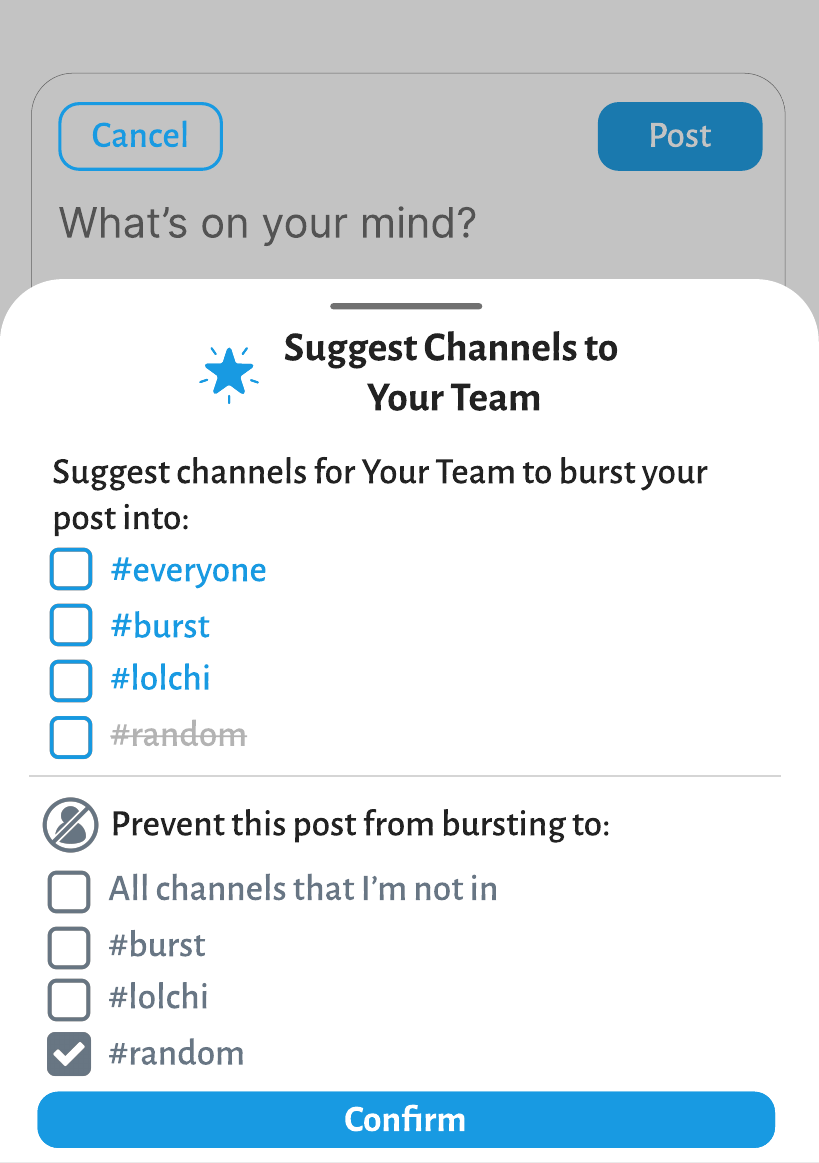}
    \caption{\textbf{Channel Suggestion and Block:} Posters can suggest channels they belong to as intended audiences to signal a post’s context and purpose. They can also block channels they wish to exclude; blocked channels cannot receive bursts.}
    \label{subfig: suggest_block}
  \end{subfigure}
  \hspace{0.010\textwidth}
  \begin{subfigure}[t]{0.28\textwidth}
    \centering
    \includegraphics[width=\textwidth]{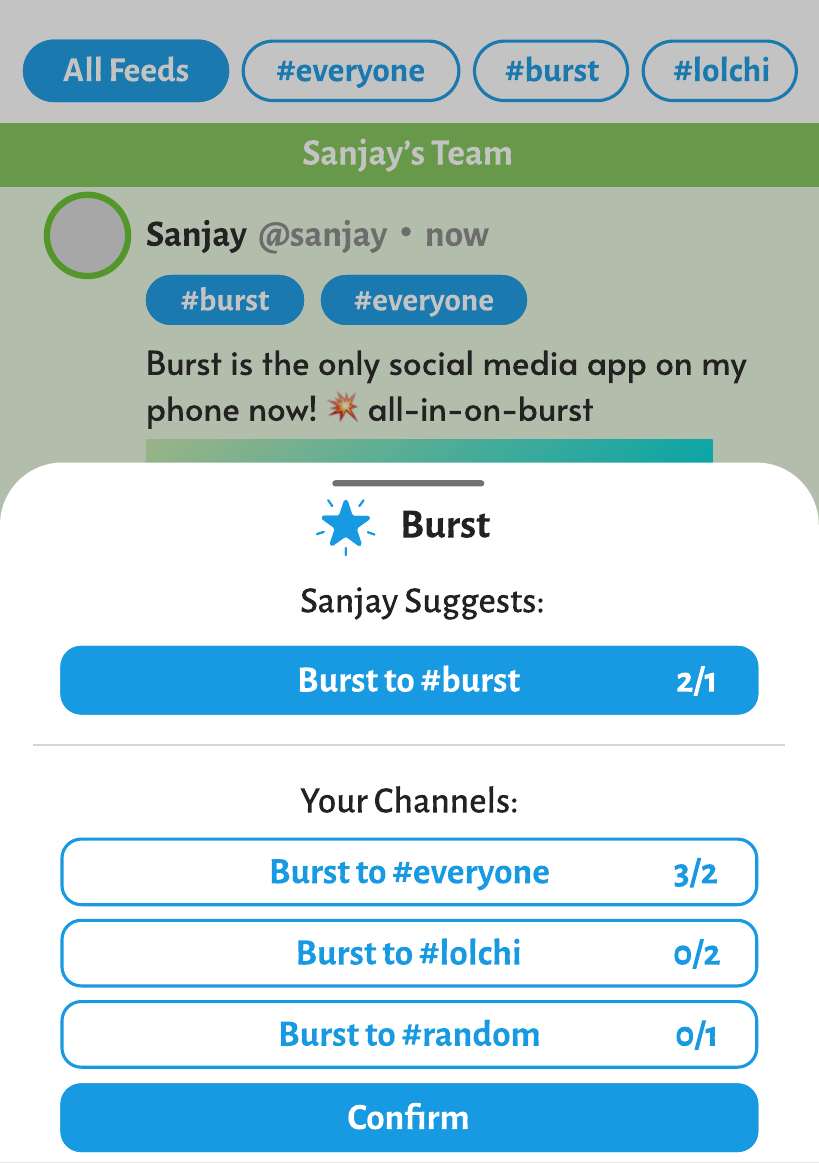}
    \caption{\textbf{Burst:} As viewers, users can vote to burst a post if they believe it is high-quality and should reach a broader audience. They can choose to burst the post to the channels suggested by the poster as well as any channels they have joined.}
    \label{subfig: burst}
  \end{subfigure}
  \hspace{0.010\textwidth}
  \begin{subfigure}[t]{0.28\textwidth}
    \centering
    \includegraphics[width=\textwidth]{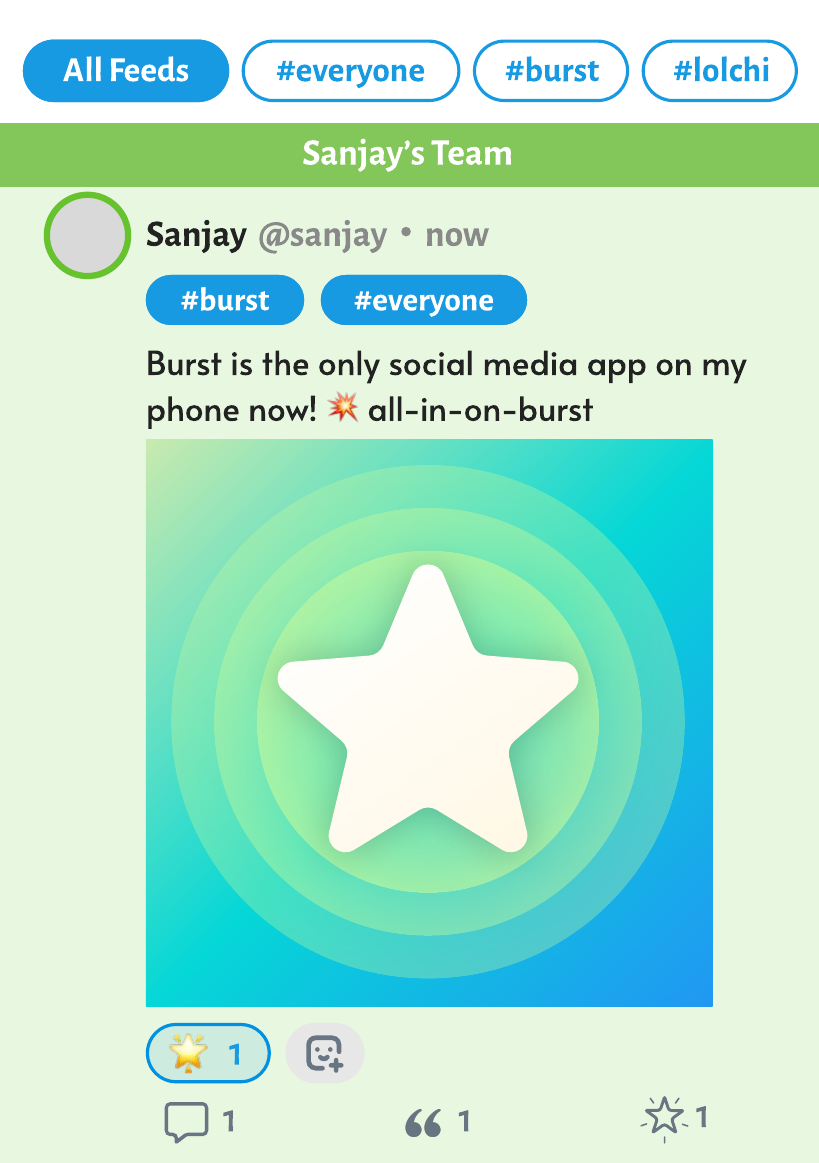}
    \caption{\textbf{Feed Page-Team Version}: The main feed shows posts burst into joined channels and posts from posters whose teams the user has joined. Posts from posters where the user is an initial reviewer are highlighted with a green background.}
    \label{subfig: burst_post}
  \end{subfigure}
  \hspace{0.010\textwidth}

  \caption{Within the Burst app, we provide the user with the capability to (a)~create and join channels, (b)~join and leave other users' teams, (c)~invite and remove users from one's own team, (d)~submit posts, with the option to suggest or block channels as bursting destinations, (e)~burst posts, and (f)~scroll the feed and check out different channels.}
  \label{fig:app-prototype}
\end{figure}

Burst features atomic text posts similar to Twitter/X, but a feed that is similar to other threaded-space social media such as Reddit (Figure~\ref{fig:app-prototype}). The feed intermixes content from all channels the user has joined, including any “Your Team” groups that the user is in, and supports standard interaction methods such as emoji reactions, replies, and quote posts. In this section, we explicate how the key elements of the \designname architecture are realized in our app.

The \appname{} app is designed to facilitate an ecosystem of different channels. Users may create a new channel by giving it a name (e.g., \#channel-name), a brief description, and a set of users (Figure~\ref{subfig: community}).
Each channel in Burst has a threshold that determines how many peer endorsements (or ``bursts") are required before a post can appear there. To reduce user burden and support a simple mental model, currently, thresholds are automatically set by the system based on the size of the target channel in our prototype system. Specifically, the threshold is proportional to the number of active members in the channel: smaller channels typically require only one or two bursts, while larger channels demand more endorsements to ensure that content has broader support before reaching a wider audience. There are also other options, such as algorithmic threshold models that dynamically adjust based on content or activity (discussed more in Section~\ref{sec: discussion}).
As shown in Figure~\ref{subfig: burst}, users can view the current burst progress for each channel (e.g., “2/3” indicates that the post has received two bursts out of the three required). This dynamic helps users understand how close a post is to being shared with a given audience and reflects the selective nature of different channels. While channels may differ in size and purpose, this thresholding mechanism provides a consistent and lightweight way to gate content flow based on community relevance and endorsement.

As users scroll through their feeds, they see content that has burst into channels that they are a member of. Posts are also tagged with other channels that they have burst into, a discovery mechanism for users to learn about and then join other new channels. To maintain simplicity for a platform currently operating at a small scale, the feed is currently sorted chronologically by post creation time, without algorithmic ranking, but it can be easily integrated with existing feed algorithms if needed (Figure~\ref{fig:burst-feed}).
Users have the freedom to join and leave any channels that align with their interests. In \appname{} currently, channels include those organized by topics, like \#hci and \#ai/ml; channels that differ in size and scope, such as the smaller \#stanford-hci and the larger \#stanford; and channels that differ in tone or norms, such as the broader \#hci vs. the humorous variant \#lolchi. We also allowed users to create channels for their own purposes. The platform provides a default \#everyone channel.

The sharing mechanism in the platform is designed to reduce individual pressure and encourage collaborative curation. Rather than relying on users to determine where and how to share their own posts with a wider audience, the system allows users who are familiar with the author or the intended recipients to vote on posts they believe should be shared with these broader audiences (Figure~\ref{subfig: burst}). Through this process, content is distributed purposefully, aiming to help it reach the most appropriate audience.
Because all content must burst into channels, Burst requires a place for content to be reviewed and curated when it is first posted and not yet in any channels. So, one distinct feature of \appname{} is that we also create a private channel for each user called "Your Team" (Figure~\ref{subfig: your_team}). To configure Your Team, each user is asked to invite a group of trusted users to serve as the first set of viewers for their posts. All posts are first sent to a user's team, who receive phone notifications that they have new content to review. As shown in Figure~\ref{subfig: burst_post}, all people who are in a user's team will see that team's content in their main feed, highlighted and separated from other posts by a green background and banner (e.g., `Sanjay's Team') to call attention to the fact that they are in the small group of users responsible for deciding whether to burst that content. Users can add or remove people from their team as they use the application. Currently, \appname{} only allows one team per user; however, in principle the user could create different teams for different kinds of content that they post. A fully alternative approach would be to allow some channels to have no burst threshold and thus allow direct posting; however, we opted for Your Team to provide more positive accountability.

Burst makes bursting content a primary action, similar in visibility to replying and emoji responses. Users may tap on the burst action underneath a post, pulling up a list of channels that they are a member of and can burst the post into (Figure~\ref{fig:app-prototype}). Each channel is accompanied with its current votes and burst threshold (e.g., a post might have one out the two bursts required to burst into \#hci). Users can select multiple channels to burst the post into if desired. If a user is a member of multiple channels where a post has burst into, they only see the post once in their feed, tagged with both channels. To reduce social pressure, bursts are currently anonymous to everyone, showing only the total burst count without revealing individual voters. We acknowledge that alternative designs might instead display burst contributors for credit, similar to retweets on Twitter/X. However, such an action would remove plausible deniability: a poster could see who left emoji reactions but chose not to burst, potentially creating friction.

Previous studies have shown that posters often have a clear imagined audience in mind when crafting their content~\cite{litt2016imagined, marwick2011tweet}. 
To support this, Burst includes a channel suggestion feature that allows posters to specify one or more channels they believe are appropriate audiences for their post (Figure~\ref{subfig: suggest_block}). These suggested channels appear at the top of the list when team members press the “burst” button (Figure~\ref{subfig: burst}), providing social cues about where the post is relevant and why it was written. This not only supports smoother content curation but also helps viewers better interpret the purpose of the post and decide whether to help it burst outward. This feature is particularly useful when posters have a specific group in mind, enabling them to direct attention without relying on unilateral broadcasting. In our current implementation, posters manually select suggested channels from the ones they have joined. Future versions could incorporate lightweight recommendation or ranking to reduce cognitive load while preserving poster control.
Conversely, posters may block the post from being burst into specific channels, in order to control their audience. If a channel is blocked, users are not given the option to burst the content into that channel. Users may also retroactively remove a post from a channel it has burst into, or block accounts if needed.

\subsubsection{Example Scenario}
Xiaoling has an idea about how Burst can help fill the gap between private group chats and public social media, and thinks it could be a good research idea for human-computer interaction.  However, she feels unsure whether others will find it relevant or valuable, so she’s hesitant to share it directly with a public audience---on the other hand, if it is a good idea, she would be happy to see it reach a broader audience. To address this, Xiaoling posts her idea to her team on \appname{}. In doing so, Xiaoling suggests the \#burst and \#stanford-hci channels as possible burst locations. 
Xiaoling’s team finds the post interesting and five of her teammates vote to burst it to \#burst, which is a smaller project channel. This far exceeds the relatively low threshold of two required for a smaller channel like \#burst, and the post is successfully bursted there. Several members of \#burst who are also members of the larger \#stanford-hci channel and think the idea would resonate with this broader HCI audience in the lab. Five of Xiaoling’s teammates also vote to burst it into \#stanford-hci, but because the \#stanford-hci channel has a higher threshold due to its larger size, the post is not immediately bursted. Later, five more members of \#burst who are also in \#stanford-hci vote for it, collectively meeting the threshold of ten for the \#stanford-hci channel. The post is then successfully bursted into \#stanford-hci, a space populated with senior researchers who react positively.
More people see the post and like the idea. Eventually, 50 people vote to burst it into \#cscw, a big channel with researchers from outside the lab. Xiaoling hadn’t joined \#cscw before, but once the post is shared there, she discovers the channel and joins it.

\section{Evaluation Method} \label{sec: evaluation}
In this section, we report a field study to understand how people utilize and adapt to Burst over a period of ten days. In this study, we triangulated the impact of Burst by analyzing (1)~behavioral data: logged actions including bursts, reactions, and comments, (2)~attitudinal data, collected through pre-study and post-study surveys, and (3)~semi-structured interviews with participants.

To understand how Burst shaped participants' social media practices, we recruited paid participants from our institution to use Burst, and observed the resulting social dynamics. This social computing systems evaluation strategy overcomes the cold start problem of populating the system and allows us to begin to understand how the system is actually used and co-opted for peoples' own use~\cite{bernstein2011trouble}.

\subsection{Participants}
We recruited participants for a ten-day field study through institutional Slack workspaces and email lists. Given that participation required users to form their own trusted teams for content curation, recruitment was limited to within the institution to increase the likelihood that participants would know and trust one another. Participants were also encouraged to invite their peers to join the study, enabling the formation of socially cohesive groups. All users---whether recruited directly or invited by others---were required to provide informed consent prior to accessing the app, allowing the research team to collect non-identifiable, aggregate in-app activity data.

In total, 55 users joined the Burst platform and provided initial consent for data collection. Among them, 36 completed both the pre-study and post-study surveys. This group provided the demographic and attitudinal data and also served as the primary sample for our behavioral analyses. Within this primary sample, 16 participants also took part in a semi-structured interview at the end of the study. All interview analyses in this paper are drawn from this subset of 16 users. Among the 36 participants, 54\% were graduate students, 40\% were undergraduates, and the rest participants were visiting researchers affiliated with the institution. In terms of academic background, 89\% were majoring in engineering, with the remaining studying statistics, design, or political science. All participants reported regular usage of at least one social media application before joining the study.

Participants who completed the pre- and post-study surveys received a $25$ Amazon gift card as compensation. Those who also participated in the follow-up interview received an additional $40$ Amazon gift card. Participants who used the app but did not complete the study-specific consent form and survey instruments were not eligible for compensation.

\subsection{Procedure}
When participants signed up for the study, they completed a pre-study survey. Participants then downloaded the Burst app to their iOS or Android devices and enabled app notifications. Following registration, they proceeded through Burst's app onboarding process within the app. This process provided a brief introduction to the main unique design features of the app, including ``Your Team'' and ``Burst''. To join the app, we required participants to invite at least three users to join their team as curators and to join at least three channels based on their interests. These requirements were established as necessary for meaningful engagement with the app: having people on Your Team ensures that participants have enough people to curate their posts, while joining multiple channels ensures that their feed contains content. At the end of the study period, participants completed a post-study survey and participated in a semi-structured interview to share their opinions and experiences using Burst.

\subsection{Data Collection and Analysis}

\paragraph{Behavioral Data}
Our analysis centers on activity measures related to cross-channel interactions. We examine the channels to which posts are burst, the poster's use of suggestions and blocking, and the alignment between the channels suggested or blocked by the poster and the actual channels to which the posts are burst. Additionally, we consider post-related actions, such as the number of comments, reactions, and bursts a post receives, as well as user-related actions, including the number of team members associated with a user and the channels these team members join.

\paragraph{Survey Measures}
Participants were asked to complete two surveys: one before joining the app and another at the end of the study period. The pre-study survey explored participants' previous experiences with other social media platforms. 
The post-study survey assessed their experiences with the primary features of Burst, such as burst, Your Team, and channels, as well as their sense of psychological safety when posting and whether their expectations and needs were met.

\paragraph{Semi-Structured Interviews}
We conducted semi-structured interviews with participants to achieve three main goals: 1)~to understand how they use and interact with the Burst feature, 2)~to understand their experiences and opinions regarding Burst, and 3)~to explore how they feel Burst influences their behavior. The interview questions focused on participants' interactions with various features, their motivations for engaging in different activities, and the impact of the design on their overall user experience. To help participants recall their experiences, participants were encouraged to open the application while responding to questions. All interviews were recorded and transcribed. 

We conducted a thematic analysis to identify key themes across our qualitative interview data \cite{braun2006using}. One author performed an initial line-by-line open-coding of the first five transcripts, generating codes both inductively from the data and in line with our research questions. Preliminary themes were developed based on recurring patterns and participant feedback, then refined as the remaining transcripts were analyzed. Coding was iteratively adjusted to incorporate new insights, with themes generated at a semantic level to capture participants' explicit statements.

\section{Results} \label{sec: results}
This section presents our findings from the field study, drawing on activity logs, survey data, and interview responses to examine the influence of the burst design.
Across the study, the 36 participants generated a total of 263 posts, including 3 quote posts and 141 replies. In addition, participants performed 297 burst actions and contributed 501 emoji reactions to posts.

\subsection{Participants burst content to curate and share content with each other}
\begin{figure}[tb]
  \centering
  \includegraphics[width=\textwidth]{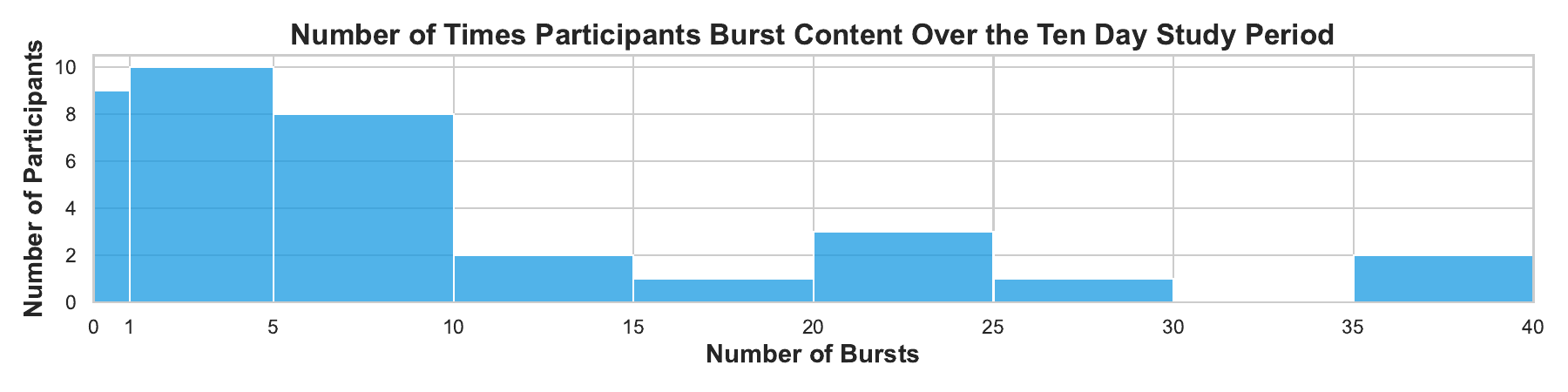}
  \caption{Participants explored the app without any requirement to use the burst feature during the study. Among the 36 participants, 27 burst at least one post. On average, participants generated a maximum of 40 bursts, with an average of 15.79 bursts per day.}
  \label{fig:burst-count}
\end{figure}

Participants actively used the burst feature. As Burst was a newly introduced feature in this app, one initial concern was that users might struggle to comprehend and use it effectively. However, log data from the app, as shown in Figure~\ref{fig:burst-count}, indicate that 75\% of participants burst at least one post during the ten-day study period. Additionally, in the post-study survey, 72.41\% participants reported considering Burst to be somewhat or very easy to understand, 68.97\% found the bursting feature to positively influence their experience, and 65.52\% enjoyed or liked the bursting feature. In the following Section~\ref{sec:lower-poster-cost}, ~\ref{sec:unique-channels}, and ~\ref{sec: form-participation} we delve deeper into how participants engaged with bursting in practice, including their experiences as posters and as viewers participating in collaborative curation.

Participants generally found bursting to be intuitive and aligned with behaviors they were already familiar with from other social media platforms. Several described it as a natural extension of how they typically shared content with friends or interest-based groups (P2, P7, P11, P12, P16). For instance, P12 mentioned:
\begin{quote}
    \textit{``Because, at least for me, when I browse social media, one of the main things I do is share with friends, right? I either share things that a friend would be interested in or share with a group of people. So, the interaction is very similar to that, which, for me, is quite typical.''} (P12)
\end{quote}

Similarly, P11 noted the familiarity of bursting to existing practices of cross-posting:
\begin{quote}
    \textit{``So, when I was using Burst, I was thinking more like if I have a server in Discord, where would I send this post if I wanted to share it with friends? For example, sometimes I'm scrolling through X, and I see a post I want to share with friends, like on Discord. So, I copy and paste the link, then go to Discord, where I have five different channels. Based on the channel tags, I decide which channel is suitable to copy and paste that link. The Burst feature was just like that. I see a post on a feed that looks like Twitter, and then I copy-paste, or 'burst,' it into a channel mainly based on the channel description.''} (P11)
\end{quote}

As expected, participants initially experienced a learning curve with bursting. However, once they became familiar with it, they engaged with it actively:
\begin{quote}
    \textit{``Since I'm not used to bursting posts in my other social media sites, it took me a couple of days to get into the habit of bursting posts I found interesting. After getting used to the process, however, it became enjoyable to burst a post to any relevant channel that I was a part of.''} (P25)
\end{quote}

While most participants responded positively to the burst feature, 6.9\% expressed some dissatisfaction due to the added consideration and moderation involved in the bursting process. For instance, P5 found the feature unintuitive and slightly burdensome:
\begin{quote}
    \textit{"I just don't like the concept, it's like making me a content moderator of the platform."} (P5)
\end{quote}
Similarly, P3 believed bursting ``introduced many layers of new considerations'' which left them ``less free to post impulsively even though the concept introduced interest and curiosity.''

\subsection{Lowering the cost of contribution: supporting safe posting and collaborative content flow} \label{sec:lower-poster-cost}

Users’ willingness to share is strongly shaped by audience context, particularly when the risks of misjudgment or reputational harm are high. Burst aims to reduce this anxiety by allowing users to first share within trusted, smaller groups and delegating audience selection to peers, who collaboratively decide whether and how a post should reach broader audiences.

Participants reported in the pre-study survey that concerns about privacy and the uncertainty of a diverse public audience were their primary reasons for hesitating to share content publicly. They expressed a preference for smaller, private spaces where they could share more openly, free from concerns about judgment or misunderstandings. For example, P25 explained:
\begin{quote}
    \textit{``I think my tone is definitely a lot more expressive when I'm in a small group since I know the people in that group won't judge me\ldots However, in a public space, I'll definitely think twice before posting something... I'll often overthink what to post and how to caption certain posts when I post publicly since I feel like I have an online persona to maintain.''} (P25)
\end{quote}

Participants described Burst's design, sitting between private and public sharing, as more psychologically comfortable. Instead of broadcasting to a large, undefined audience, users could post within their team and rely on their peers to determine the appropriate reach. For example, P10 mentioned:
\begin{quote}
    \textit{``I think it probably will be in between but a little bit closer to like my personal group chats, because my teams are a little bit smaller, so I don't feel as burdened by hundreds of people seeing my post at once. So I'll probably feel more comfortable posting often compared to posting to the public.''} (P10)
\end{quote}

By initiating sharing in a trusted group, Burst allowed participants to focus on expressing their ideas rather than managing their audience. The peer-curation mechanism helped alleviate the cognitive load and hesitation associated with posting:
\begin{quote}
    \textit{``It actually reduces the overthinking process before posting. To be honest, it's like, if you're trying to share something you find interesting, you just share it. If it's actually interesting, then people will burst it to other channels so more people can see it. If it turns out not to be that interesting, people will just see it and react or reply to you, and that would be more private.''} (P7)
\end{quote}

``Your Team'' helped address some of these privacy concerns by enabling participants to feel more confident sharing content with a broader audience, like for P12:
\begin{quote}
    \textit{``Being able to have things reviewed by your smaller team before they’re shared out with the world is quite beneficial, in terms of limiting the barrier or fear of posting. Things do get cross-posted, but for them to be cross-posted, they have to be posted for the first time. The first interaction—like the initial burst—helps overcome that activation energy to post for the first time. Once something’s been posted, now there's so much more content to be shared. Even though the burst interaction is similar to a cross-post, it actually enables more original posts.''}
\end{quote}
P1 described trusting their team to only burst content suitable for a broader audience:
\begin{quote}
    \textit{``I guess once it gets outside of my team---I guess other people are then able to burst the other channels, right? I guess that part is where it feels like I have less control. But at the same time\ldots things that make it out of my like [team]\ldots are usually, like, fine things for everyone else to see, if that makes sense. It's like they already do a good quality control on my content, so I feel pretty good about it.''} (P1)
\end{quote}
Similarly, P10 considered the bursting feature as \textit{``kind of like validation, that someone else also likes your post and also wants to share the post.''}

The burst feature was largely successful in helping posters reach their desired audiences, and the majority of posts burst to at least one group. Only 36\% stayed solely within the poster's team. A majority of participants, 63\%, were satisfied with their team’s decisions on which channels to burst their posts into, and 47\% of participants felt that the actual audience matched their intended audience to some degree. The channel suggestion feature helped posters orient toward channels suited to their intended audience, with 63\% of participants reporting that suggestions made them more comfortable when posting, and 53\% that the ability to block specific channels increased their comfort level. P1’s feedback echoes this sentiment:
\begin{quote}
    \textit{``I felt like I had decent control over the audience for each of my posts since my team usually bursted the content into channels I had suggested. This allowed me to have control over who could see my content.''} (P1)
\end{quote}

The collaborative approach of bursting, where team members helped decide where posts would be best received, also helped posts sometimes reach audiences that the original poster had not considered or had access to. 
P12 appreciated this broader reach:
\begin{quote}
    \textit{``I wasn't part of that channel, so if I had made the post myself, I wouldn’t have been able to reach that channel. But when it was reposted or burst by someone else, it was shared into that channel, which I think was the most appropriate one for it.''} (P12)
\end{quote}
Similarly, P12 was pleased to reach an unexpected but welcome new audience:
\begin{quote}
    \textit{``I think I would be happy because it’s like when it gives you something you didn’t know you wanted. Giving people something they didn’t realize they wanted is really valuable, especially if it’s something I hadn’t thought of myself. And if I wasn’t happy with it, I could always block those channels, which is a useful feature.''} (P12)
\end{quote}

The bursting process helped provide additional context for posts, reducing the risk of misinterpretation. For example, P7 noted that the choice of a specific channel \textit{``provides more context and background knowledge to help interpret [their] post.''}
This added context proved particularly valuable in ambiguous situations. P19 shared an example where a post could be interpreted in multiple ways:
\begin{quote}
    \textit{``Sometimes, it’s hard to tell if something is meant to be funny or not---it’s just kind of ambiguous. For example, I posted a picture of my roommate with a horse head really close to the camera. I thought it was funny, but it could also be seen as cute or just a slice-of-life moment, sharing about my roommates or our weekend.''} (P19)
\end{quote}

Still, some participants expressed dissatisfaction with the bursting feature, preferring to have more control over selecting their audience rather than relying on peers to filter posts. As P22 mentioned, \textit{``I kind of don't like how it motivates a peer-based filter on posts. I'd rather be able to choose what I click on myself from the full sample of people''} Another participant echoed and emphasized the preference for personally choosing the audience groups rather than using the group-based approach:
\begin{quote}
    \textit{``I’d definitely trust them more in a real setting if closer friends were involved. But honestly, I’d much rather approach someone myself than risk offending the group by doing it formally''} (P5)
\end{quote}
This preference for individual control over audience selection was especially high when posts contained sensitive or personal content:
\begin{quote}
    \textit{``If I post something that is very personal or something like that, then I don’t even trust myself sometimes to channel it to the right audience. So I certainly don’t trust my friends with that, even my closest friends''} (P11)
\end{quote}
Overall, we find that the multi-layer sharing structure reduced users’ hesitation to share content. The ‘Your Team’ channel provided a psychologically safe starting point, allowing participants to test ideas with trusted peers before those ideas reached larger audiences. This layered visibility reassured participants and supported more authentic content sharing.

\subsection{Bursting connects groups while maintaining their unique character} \label{sec:unique-channels}

\begin{table}[!htbp]
    \centering
    \small
        \begin{tabular}{|p{1.8cm}|c|c|p{1.2cm}|p{5.5cm}|}
        \hline
        \textbf{Channel Name} & \textbf{Members} & \textbf{Participants} & \textbf{Bursted Posts} & \textbf{Description} \\
        \hline
        \#everyone & 95 & 36 & 7 & The default channel, automatically joined by all users. \\
        \hline
        \#burst & 64 & 21 & 4 & A channel for discussions about the app and its latest updates. \\
        \hline
        \#lolchi & 54 & 14 & 2 & A channel for sharing memes and humorous content related to HCI. \\
        \hline
        \#stanford-hci & 53 & 19 & 4 & A channel for HCI discussions, primarily involving students and faculty from Stanford University. \\
        \hline
        \#stanford & 33 & 21 & 52 & A channel covering topics and events related to the institute. \\
        \hline
        \#lol & 29 & 9 & 4 & A channel for sharing memes and humorous content. \\
        \hline
        \#random & 28 & 15 & 20 & A channel for off-topic discussions and casual conversations. \\
        \hline
        \#ai/ml & 27 & 18 & 31 & A channel for discussions about AI and ML. \\
        \hline
        \#foodie & 25 & 17 & 14 & A channel focused on food, including sharing recipes and recommendations. \\
        \hline
        \#gym* & 25 & 18 & 38 & A channel for discussions on fitness, workout tips, and gym-related topics. \\
        \hline
        \#rollingtopics & 23 & 3 & 1 & A channel for academic discussions on research and papers. \\
        \hline
        \#hci & 22 & 5 & 9 & A channel for general discussions focused on HCI. \\
        \hline
        \#hot-take & 21 & 12 & 10 & A channel for sharing provocative or controversial opinions. \\
        \hline
        \#students-msb & 21 & 8 & 7 & A channel dedicated to students working in an HCI lab at Stanford University. \\
        \hline
        \#gto* & 16 & 9 & 44 & A channel for members of a club named ``game theory optimal''. \\
        \hline
        \#forum-pets & 15 & 7 & 11 & A channel for discussions related to pets and pet care. \\
        \hline
        \#music & 14 & 10 & 9 & A channel focused on music-related discussions. \\
        \hline
        \#event & 13 & 5 & 6 & A channel for sharing information about local or recent events. \\
        \hline
        \#festive* & 7 & 2 & 0 & A channel for discussing festivities. \\
        \hline
        \#locked-in* & 6 & 5 & 7 & A channel for discussing focused work by students and researchers. \\
        \hline
    \end{tabular}
    \caption{There are twenty channels in which participants were primarily involved, and four of them are created by participants. Since both study participants and non-participant users coexisted within the app, the column `Members' represents the total number of users in each channel, while `Participants' indicates the number of study participants in that channel. The channels cover a variety of topics, with some intentionally exhibiting overlapping or hierarchical relationships. For example, based on the scope of the channel's topic, \#stanford > \#stanford-hci > \#students-msb, suggests that these channels may contain different types of audiences. (* indicates that the channels were created by the participants.)}
    \label{tab:channels_participants}
\end{table}

\begin{figure}[tb]
  \centering
  \includegraphics[width=\textwidth]{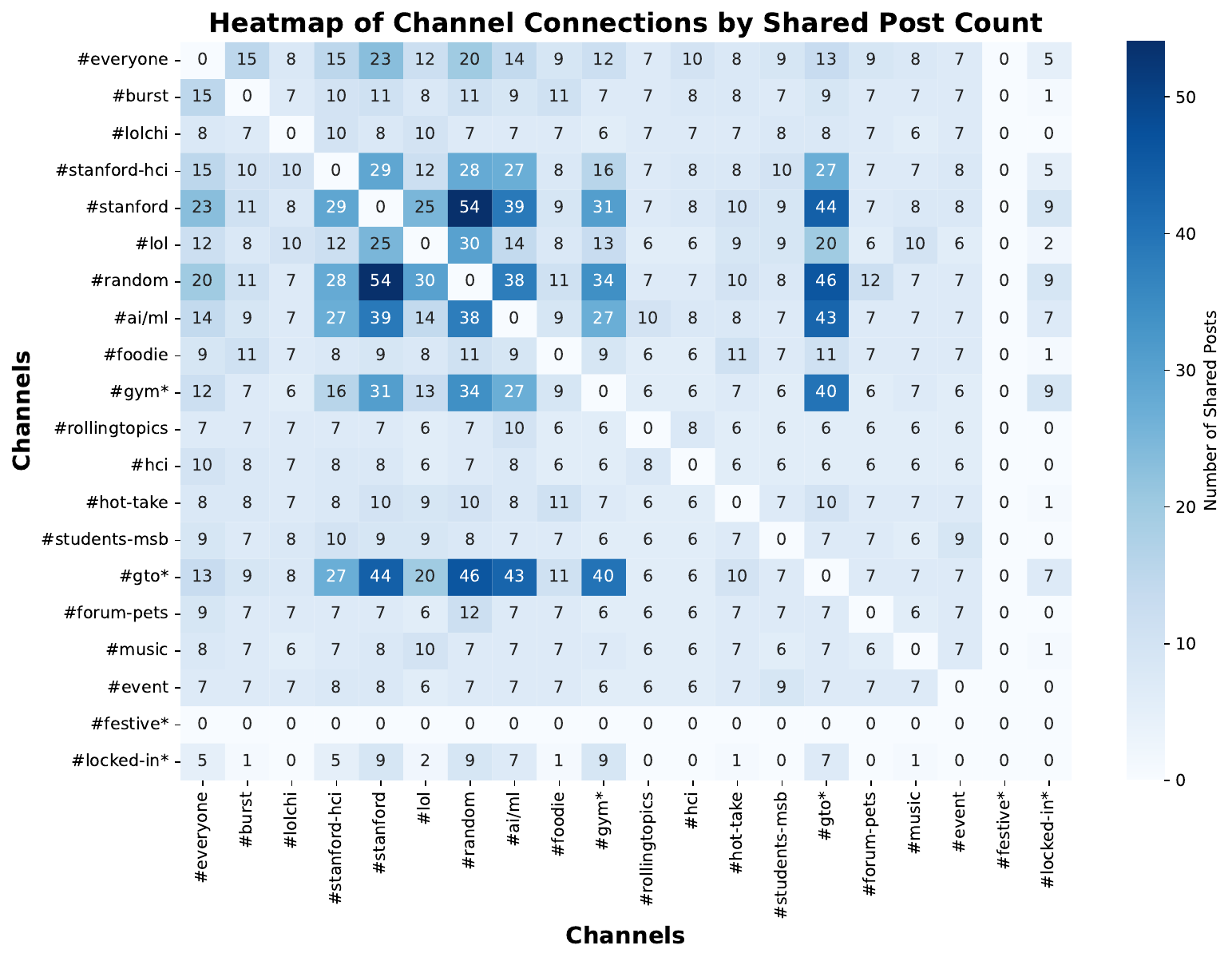}
  \caption{Channel connections, constructed by burst across multiple channels, are represented here by the number of overlapping posts. This heatmap shows post overlap across all twenty channels, ordered from largest (top/left) to smallest (bottom/right), and * marks channels created by participants. We observe significant overlap between broad channels, like \#random and \#stanford with more niche channels.}
  \label{fig:burst_heatmap}
\end{figure}

As shown in Table \ref{tab:channels_participants}, the Burst platform contained a range of channels, from large public spaces like \#everyone to smaller, specialized channels such as \#event and \#festive. Through the bursting feature, posts shared within one channel could propagate across others. Survey data highlighted user preferences for this function: 79\% of participants felt that posts should be shared across multiple channels at least some of the time.

Figure \ref{fig:burst_heatmap} illustrates some patterns and clusters in these connections. Smaller channels like \#gto maintained notable connections with larger, more general channels such as \#gym, \#random, and \#stanford, while larger channels, like \#stanford-hci and \#stanford, shared a substantial number of bursts, with 26 posts cross-shared between them, indicating strong linkages between these larger groups. However, there are fewer connections between smaller channels, likely due to the focused nature of these groups, in contrast to the broad mandate of the larger channels.

Participants expressed perceptions of varying connection strengths across channels:
\begin{quote}
    \textit{``There are some channels that tend to be grouped together. I feel like posts would often be shared concurrently across certain channels together. There are some cross-posting, and maybe overlapping conversations happening. But then there are other channels that did feel quite separate from others.''} (P1)
\end{quote}

We also noticed various sharing patterns: 36\% of posts were shared only within teams, 18\% were bursted once, 16\% of posts moved from smaller channels to larger ones, and 30\% of posts were shared from larger channels to smaller ones. Participants reported different expectations for bursting content depending on the direction of the content flow. For bursts from smaller to larger channels, many participants (P7, P8, P14, P19, P28, P30) saw this as an opportunity to reach a broader audience. For instance, P8  considered bursting to a larger audience as a way to share popular opinions and promote content they wish to disseminate widely, with P8 explaining: 
\begin{quote}
    \textit{``If it’s a popular opinion or an idea that everyone shares but hasn’t fully considered, I think it’s important to burst it. That way, it can be shared more widely, allowing more people in different channels to see it.''} (P8)
\end{quote}

For bursting from larger to smaller channels, participants mentioned that they viewed this process as a moderating function, selecting the most relevant and valuable posts from the broader contents in the larger channel:
\begin{quote}
    \textit{``Someone has gone through and done some curation from another channel. Maybe there's a general ML Papers channel that everyone is posting to, and then someone creates their own sub-channel, where they post what they think is the best content from that larger channel. I’d probably be more likely to follow the one curated by a person or group, rather than the big one where everyone is posting.''} (P28)
\end{quote}

Likely due to the curation efforts of our participants, both classes of group---small/private and large/public---provided unique experiences despite being connected. Participants saw smaller, more specific channels as spaces that fostered personal interactions, similar to direct messages or personal channels. For example, P11 described how the two classes of groups differed:
\begin{quote}
    \textit{``I feel that posting in more generic channels is like retweeting while posting in more specific channels is more like sending a direct message. I have a general feed of content from different social media, and I re-burst it into personal channels. The reward from that is knowing that a friend actually reads what I sent, and that feels rewarding.''} (P11)
\end{quote}

Many participants noted that ``Your Team,'' a uniquely private channel, contributed to a sense of intimacy on the platform, making it feel like a space for close-knit interaction (P2, P4, P7, P8, P14, P16, P17, P18, P20, P22, P23, P25). 47\% of participants reported choosing not to burst certain content because they felt it was ``only relevant for our small group or the poster's team,'' suggesting that team members regarded "Your Team" as a space for more in-group content and would not further share with broader outside audience.

P7 shared, ``Your team is your smaller private group'' adding that they were comfortable with content remaining within this team since it still supported interaction among close connections.

\subsubsection{Connected groups facilitate user self-expression across contexts}

The majority of participants noticed a difference in the content or atmosphere across the twenty channels on the Burst platform.
Due to the diversity of channels, participants felt better able to express their different interests and affinities without having to maintain boundaries via alternate accounts (P1, P7, P8, P14, P19, P28). The channel-based structure enabled users to post varied content intended for different relevant groups, easing self-presentation concerns common on more generalized platforms.

For instance, P28 mentioned that the various channels could allow them to post more different topics about themselves compared to other broad audience social media:
\begin{quote}
    \textit{``I used to use Twitter personally, but since Twitter is such a big part of academia now, I stopped using it in a personal capacity. I thought, `Oh, this is really more of a professional account now.' I used to post more lighthearted, personal things, but once I became a PhD student and started using Twitter in a more professional way, I stopped posting that kind of content. A big part of it is that most of my followers now are professionals, and I don’t want to clutter their feeds with personal posts that probably don’t interest them. I’d prefer having one account where I could post a variety of content and share it across different channels.''} (P28)
\end{quote}
Similarly, P1 mentioned that joining multiple channels is similar to the strategy they used by having multiple accounts for the content segmentation they have on other social media. Besides, the bursting mechanism also facilitated the sharing between different audience groups. 
\begin{quote}
    \textit{``The community setup here is most similar to Slack, which I’m familiar with, rather than Twitter or Facebook. I don’t really use Facebook, and on Twitter and Instagram, my main feed is much more all-encompassing, which brings the risk of context collapse. Here, everything is nicely segmented. In a way, I've tried to replicate this on Instagram by creating four different accounts: one for personal life, one for photography, one for food, and one for design. But I can’t easily move content across these different accounts. In that sense, having something like Burst would be helpful. 
    ''} (P1)
\end{quote}
Because content on Burst is only visible within the channels it is burst to, viewers primarily see posts aligned with their shared interests—reducing the pressure on users to manage how they are perceived by unrelated, uninterested or disapproving audiences. This segmentation minimizes unintended exposure across social contexts and helps prevent context collapse. By being exposed only to content relevant to a shared interest, users engage with a limited slice of the poster’s identity rather than their entire posting history, further easing the burden on posters to manage multiple social roles.

However, due to the small size of some channels, the threshold for bursting content to these channels is often set at one, giving individual users considerable influence in the curation process. This low threshold can lead to content being burst into inappropriate channels. For instance, P7 noted that \textit{``sometimes random people randomly burst the post out.''}

The high overlap in membership across channels led some participants to view these channels as similar, often resulting in posts being burst to multiple channels even if they were intended to cover distinct topics. This overlap could blur the distinctions between channels (P9, P14, P15, P23), as voiced by P14:
\begin{quote}
    \textit{"For example if I have a channel called \#gym and \#gto and I know the same set of people are in both the groups, I might burst it to both the channels."} (P14)
\end{quote}

Furthermore, due to the limited number of users on the platform, participants mentioned that there were too few channels (P25, P8) and insufficient content (P19, P22),  to foster a strong sense of community. Since most participants were from the HCI field and affiliated with stanford, the available channels primarily focused on these areas, potentially limiting the diversity of topics and perspectives (P16, P18, P25).

\subsection{Bursting serves as a meaningful form of participation} \label{sec: form-participation}

Bursting served as a lightweight, low-effort interaction that allowed participants to contribute without the extensive effort required for quoting or replying, while offering a higher level of engagement than simple reactions. As shown in Table \ref{tab:activity-distribution}, bursting occurred less frequently than reactions but was nearly twice as frequent than quoting or replying. 

\begin{table}[h!]
\centering
\begin{tabular}{lccccc}
\toprule
 & \textbf{Create} & \textbf{Reply} & \textbf{Quote} & \textbf{React} & \textbf{Burst} \\
\hline
\textbf{Mean} & 3.25 & 3.81 & 0.08 & 13.69 & \textbf{8.14} \\
\textbf{Std Dev} & 5.09 & 7.86 & 0.36 & 12.33 & \textbf{10.41} \\
\textbf{Median} & 1.0 & 1.5 & 0.0 & 10.5 & \textbf{4.0} \\
\bottomrule
\end{tabular}
\caption{Bursting, which participants did an average eight times during the ten day study, observed levels of interaction in between posting (an average of three times) and emoji reacting (an average of thirteen times). Bursting was seen as an intermediate level of signaling commitment between these two actions.}
\label{tab:activity-distribution}
\end{table}

User feedback further reflects positive perceptions of the burst feature. Both reacting and bursting were rated favorably, generally between ``somewhat liked'' and ``liked'' on a five-point Likert scale. In interviews and surveys, participants described bursting as easy and enjoyable. For example, P1 noted, \textit{``Bursting was a very easy and lightweight interaction''} while P14 added, \textit{``It's really easy to burst content on Burst. I like how it gamifies it while still trying to make the post available to the right audience.''}

While being relatively lightweight, bursting provided a stronger sense of contribution than simple reactions. It allowed participants to feel more involved in supporting their friends’ content, as P1 explained:
\begin{quote}
    \textit{``It feels kind of like your friend is reaching out, and they're asking for your advice, and I think that always---I don't know---it feels nice to be needed. So I was always okay with reviewing my friends' posts, and I'd always do it pretty soon if I had the time.''} (P1)
\end{quote}

For several participants, bursting therefore felt like a responsibility, especially when they were part of a friend’s close team. They mentioned that being motivated to burst posts was mainly due to a desire to help their friends' content reach a broader audience (P5, P7, P14, P16, P22). As P16 explained, \textit{``If I don't share it, my friends' posts might not be viewed by others, so I need to share them.''} Similarly, P7 expressed that bursting felt purposeful, involving a sense of responsibility to ensure the post reached a relevant audience:
\begin{quote}
    \textit{``For bursting, the feeling is more like, `Okay, you feel it’s your responsibility to share it purposefully.' So, it’s like your real friend is not only sharing their post with you, but also wants to make sure that, as a responsible user, you’re sharing it with an audience that will find it relevant, not totally irrelevant. When there’s no reaction or burst, you probably think of bursting first because, rather than just reacting to the post, you want to ensure there’s an audience for it. I do feel the responsibility and have the power to choose whether to burst it or not. That brings a brand new experience—it’s something totally new. You’re helping decide who might see it or even if anyone needs to see it.''} (P7)
\end{quote}

Some participants also saw bursting as a form of reciprocal exchange, helping to share others’ posts with the hope that their own content would receive similar attention. P5, for example, described this mutual aspect of bursting:
\begin{quote}
    \textit{``I want my posts to be shared in other channels, and I’ve helped share other people’s posts to channels as well. Yeah, so I did that. I think it was kind of a quick exchange---I’d help post theirs, hoping they’d help post mine, honestly. I would want my post to go into the broadest possible channel because I wasn’t sharing anything specifically; I just wanted it to reach a big channel. Likewise, I generally used to post other people’s content onto larger channels as well.''} (P5)
\end{quote}

Participants felt responsible for bursting not just toward the poster, but also toward other app users. Since bursting influenced the content of the feed, they took responsibility for reading and evaluating posts before deciding whether to burst them to produce a better experience for other users. This sense of responsibility was particularly evident among low-posting participants, who did not frequently create content themselves. Rather than passively browsing the feed, they became more attentive and thoughtful, treating the bursting process as a meaningful responsibility:

\begin{quote}
    \textit{``It felt like the things I was doing---like bursting and engaging with posts---had a bigger impact. It definitely felt more involved, like I was really there.''} (P2)
\end{quote}

At the same time, participants felt the responsibility of \textit{not} sharing content that they found undeserving.
For instance, P7 mentioned that they refrained from bursting content deemed too private, random, or suspicious, and P1 highlighted a similar moderation-like approach to bursting:
\begin{quote}
    \textit{``I didn't burst that one because I thought it was kind of spammy. I was like, oh, this is, I don't know, maybe it would be bad for the community to have this post out there. I don’t know what 'bonking' is—does it mean trolling or making fun of someone? So I decided not to do anything about that one. It felt like a form of quality control, both for the larger community and also considering whether the person posting might feel embarrassed if this content is shared.''} (P1)
\end{quote}

As a whole, participants noted that these curatorial bursting efforts improved the overall quality of the feed, ensuring it remained relevant and meaningful, especially in the broader, more public spaces. The bursting mechanism served as a filter, preventing lower-quality or promotional content from being widely shared:
\begin{quote}
    \textit{``An objectively well-curated feed would be ideal. But if it’s the other kind of posts—like publicity, promotions, junk, or spam—a small group of people pushing it might still burst it, but the general public wouldn’t. This means it wouldn’t meet the higher thresholds, which I think would be a good thing.''} (P11)
\end{quote}

However, not all participants viewed the burst feature as essential or unique. For example, P22 remarked that bursting felt too similar to upvotes on other platforms. And P2 questioned both the uniqueness and the need of the bursting feature:
\begin{quote}
    \textit{``I don't see how this is significantly different from other social media platforms like Instagram. I understand there are different channels to follow and it's more customized, but I don't know if people need the content they share and consume to be shared in such categorized ways''} (P22)
\end{quote}

\section{Discussion} \label{sec: discussion}
In this paper, we have articulated a social media model where curation, or ``bursting'', is a central design tenet. We observed in our initial field evaluation that this model, as manifested in the Burst application, resulted in users posting and routing content to destinations both narrow (`\#forum-pets', `\#lolchi') and broad (`\#everyone'). In this section, we reflect on the consequences of this design, and possible future directions.

All evaluation methodologies make tradeoffs, and this one is no exception. Most salient is that our evaluation focused on usage over approximately one week, and with a relatively small community of a few tens of people. These limitations are unsurprising for a field test of a research social computing platform and are not disqualifying~\cite{bernstein2011trouble}, but need to be contextualized. For example, one of the claims of this work is that burst-style mechanisms can help facilitate participation at different scales, but we do not yet have a truly large-scale socio-technical system for validation. There was activity in narrow-purpose communities~\cite{hwang2021people}, for example posts to a channel for students in a particular research group, but we do not yet know what might happen if the platform grew to thousands or millions of people. Likewise, while social computing systems research need not establish steady state usage with so many people~\cite{bernstein2011trouble}, we need to be careful not to overextend the conclusions of this smaller-scale evaluation. 
One potential scalability challenge is the increased mental effort required of navigating a large number of channels. As users participate in more channels, they would struggle to identify appropriate channels for bursting and suggesting. To address this, future large-scale implementations could incorporate algorithmic recommendations to help users identify appropriate channels, streamlining the decision-making process and reducing cognitive load.

The smaller scale of this study also meant that the platform did not have to deal with antisocial behavior that we expect to see in larger social spaces. Participants did agree to a code of conduct as part of their IRB agreement upon joining the platform, and the research team was present and able to step in if needed. However, because we did not observe any publicly bursted antisocial content, this was not necessary. Your Team acts as a first filter for such behavior, though it certainly can be routed around: like any socio-technical system, no technical feature can make strong guarantees, nor should it be expected to. Later in this section, we will reflect on how the burst model should adapt if the platform were to expand, which necessarily would bring along moderation concerns and politics.

A third limitation of our study is that everything was ``close to home'' in the sense that our participants all shared thick offline context: while few were actually direct friends, they all enrolled at the same university, and many of them were the same major, and members of “Your Team” were often composed of individuals who already knew each other. Of course, in many traditional social networks, this is true as well: we often connect with people with whom we share offline connections~\cite{boyd2008youth} and even in online communities, close relationships are more likely to emerge from pre-existing connections~\cite{boyd2007social}.
However, we did not, for example, have participants representing two different universities, dramatically different cultural contexts, or fully online communities. This cultural homogeneity meant that we were much less likely to see conflicts due to cultural or norm differences. However, it also meant that the community was more able to overcome a cultural cold start problem due to collective social capital~\cite{kokkalis2017founder} applying swift trust~\cite{meyerson1996swift} to others at the university.

Circumscribed by these limitations, it is worth reflecting on what we have learned about bursting as a basic design concept in social media. Cross-posting has been extant for at least as long as Usenet, but it is typically seen as a necessary evil (e.g., ``[Apologies for cross-posting] at the top of an email'') rather than a desired core element. However, those who consume content online are also those most motivated to share it~\cite{bernstein2010enhancing}: Burst taps into this mechanism. We observed that curating a friend's post with a burst is a way to demonstrate care, both to the person who posted it and to the people in the channel that would receive the burst: in this way, bursting is not unlike creating a mix tape. In terms of levels of thought and care, bursting was above likes and favorites, but below quote posting or replying---opening up a middle ground of contribution.

In theory, Burst could exist without the initial Your Team community: you could directly post to a set of channels where your content is already trusted or above-bar. In other words, Burst doesn't explicitly entail or require having a team. However, in the spirit of ``getting the design right''~\cite{buxton2010sketching}, we found that Your Team helps address several important experience details. First, it creates a safe space for experimentation: users can try out posts and trust that their closest friends can pocket veto the content if it's not worth sharing. Second, it creates an important interaction hook: because the application notifies you to review a new post shared with a team that you're on, it provides lower-activity users with something prosocial to do without them necessarily needing to create content themselves. In other words, it meant that users had things to do on the platform other than posting, which they may or may not want to do. One other question about the design of Your Team that remains unanswered: should there be a fixed limit on the size of a user's team? If the user wants their unfiltered thoughts to go to 1,000 people, should that be possible? Our intuition is that team size ought to be limited (e.g., to fifty people), to a size large enough that it doesn't put negative pressure on people to make hard decisions between close friends but also forces some decisions and avoids just recreating a traditional network-based~\cite{zhang2024form} social media site with Your Team as followers.

At a more conceptual level, bursting draws inspiration from urban planning insights such as Christopher Alexander's \textit{A Pattern Language}~\cite{alexander1977pattern}, which is known most famously in human-computer interaction for introducing the idea of design patterns~\cite{landay2003design}. Most of the focus on these design patterns has been on what Alexander might call Building Patterns rather than Town Patterns, meaning that they focused on more micro-scale design goals than meso- and macro-scale socio-technical goals. In Alexander's terms, most of our social media designs today that take a global town square metaphor are fundamentally flawed:
\begin{quote}
    \textit{``People are mixed together, irrespective of their lifestyle or culture. This seems rich. Actually it dampens all significant variety, arrests most of the possibilities for differentiation, and encourages conformity. It tends to reduce all life styles to a common denominator.''}~\cite{alexander1977pattern}
\end{quote}
With this work, we hope to expand discussions of ways that social computing designs can facilitate a broader range of spaces between small groups~\cite{hwang2021people} and large public squares.

Bursting, as a design strategy, could be integrated into other social computing designs besides a Twitter-style post network. Burst really only requires discrete communities (spaces)~\cite{zhang2024form}, as well as a mechanism for sharing content between them. For example, our early prototypes of Burst operated as a Slack application. Channels in Slack served as the spaces, and a special Burst reaction emoji allowed people to vote to Burst content with other channels. When a Slack post received two or more  Burst reactions, the Slack application burst the content to the other channel. We used this application for several months both internally within our own Slack workspace, as well as in a shared Slack channel between our research group and two research groups at other universities. In this way, group members could burst interesting content in their own internal Slack channels to a broader community of social computing researchers. 

We find it important to reflect on the risks to society posed by the introduction of bursting or burst-based social media. One clear risk is that bursting can be coopted by motivated actors, astroturfed, or otherwise manipulated to burst content. The system could potentially be exploited by bad actors using bots or multiple accounts to manipulate bursts and disseminate spam or propaganda. For example, a group of political partisans might agree to all burst each others' content over the vote thresholds to get all their posts into a global \#everyone channel. For simplicity of users' mental models and implementation for this pilot release, we leave Burst open to this kind of manipulation. However, as the platform grows, the proper response is to integrate protections against this kind of behavior into the Burst thresholds. The natural algorithmic bulwark would be a variant of the Cura algorithm~\cite{he2023cura}, which would estimate the responses of the diverse members of a destination channel. For example, if three people all tried to burst a piece of partisan content into \#everyone, the Cura algorithm would recognize that those three users have highly correlated opinions and would not ``count'' their bursts much more than had only one of them voted---it would require support from a broader segment of the community before allowing the content to burst into the channel. Successful implementation of this approach would benefit from the creation of backstage spaces for each channel, as in the Cura system, where interested members of each channel could peruse content that had some bursted support but was not yet over the burst threshold. With this approach in place, it would become successively more difficult to push content into a channel the broader and more wide-ranging its active audience is.

Another risk is unwanted publicity. Burst allows users to restrict the channels that their posts can go into, and also allows them to withdraw a post from any channel that it has burst into. However, there is always a risk that a user might not use this feature, then overlook the context collapse that would arise when the content bursts out of their local channels and into a wider community. For example, a user might post a political opinion to their trusted friends, but if it gets shared more widely, someone might forward it to others who share their own views rather than the poster's views as an expression of moral outrage~\cite{brady2021social}. One possible mitigation here is defaulting all users into a consent-based~\cite{im2021yes}, opt-in model for bursting: when their post reaches the threshold to burst into a channel, they receive a notification and must affirmatively agree to let the post burst. A middle ground might involve allowing bursting into channels that the user is a member of, but requiring consent for channels that the user is not a member of.

One additional risk is whether Burst might facilitate the creation of echo chambers online. It should first be noted that there is not yet strong empirical evidence that social media echo chambers actually \textit{cause} polarization or other antisocial outcomes~\cite{tornberg2022digital}, and indeed most social media users are exposed to \textit{more} cross-cutting news than those who do not use social media~\cite{fletcher2017news}. However, on the assumption that this might still be true, let us take the risk at face value. For example, compared to other social networks, would bursting amplify self-selection into like-minded channels, or suppress countervailing views? This could come out in two directions. If Burst creates a healthy ecosystem of medium-sized communities, it might push people out of narrow spaces, or create meso-level communities with more medium levels of disagreement rather than highly polarized platform-level disagreement. However, if people cluster either into very small or very large groups, this intervention might not succeed. 
Compounding this concern is the potential for channel saturation, where frequent bursting leads to homogenization across channels, eroding their unique character. This risk would likely be amplified by the relatively homogeneous participant pool in our study, drawn from a single institution with shared backgrounds. However, even within this setting, we observed notable differentiation among channels, suggesting that structural features of the system can still support channel diversity. Future deployments with more diverse user communities could further illuminate how channel distinctiveness evolves at scale. Implementing adaptive burst thresholds based on channel size, content type, or user diversity could help preserve the unique identity of individual communities while mitigating the formation of echo chambers.

Finally, to mitigate potential harms and uphold basic rights, it is important to maintain platform-level community guidelines, as commonly employed by platforms like Reddit and Discord, to safeguard against harmful behaviors and promote inclusive participation.

\section{Conclusion} \label{sec: conclusion}
In this paper, we introduced Burst as an innovative design approach to social media, instantiated in the \appname{} system. Burst navigates a middle ground between the paradigms of small group interactions and large public sharing by fostering connections across typically isolated groups. The design specifically emphasizes bridging diverse groups and the broader public through collaborative content curation, enabling dynamic interactions across various communities and different audiences.
A ten-day field study demonstrated that bursting is an engaging feature that supports content sharing and builds connections between different audience groups. It also supports a stronger sense of safety in content sharing and more effective audience targeting. Guided by our evaluation and design process, we reflect on the consequence of Burst design and offer suggestions for adapting the Burst strategy to other social computing designs in the future, aiming to enhance content sharing and foster better social media.

\begin{acks}
We thank our anonymous reviewers for their thoughtful feedback and constructive suggestions. 
We also thank the many individuals who contributed to this project at various stages, including Brian Park, Carmel Limcaoco, Emily Deng, Emma Wang, Estella Zhou, Jatin Gajera, Kris Jeong, Lokesh Bansal, Mitchell Gordon, Nicole Garcia, Pauline Arnoud, Ryan Li, Star Doby, Tianqi Liu, and Wanrong He. We are grateful to Carolyn Zou, Dora Zhao, Farnaz Jahanbakhsh, and Helena Vasconcelos for their assistance during the early prototyping phase. Their ideas, feedback, and efforts have been invaluable throughout the development of this work. 
This research was supported by the Stanford Institute for Human-Centered Artificial Intelligence, the Office of Naval Research, and NSF Grant CCF-1918656.
Lindsay Popowski was supported by a Stanford Interdisciplinary Graduate Fellowship and a Google PhD Fellowship. Michael Bernstein wishes to acknowledge the Center for Advanced Study in the Behavioral Sciences (CASBS) for providing an engaging intellectual home for developing these ideas.
\end{acks}

\bibliographystyle{ACM-Reference-Format}
\bibliography{main}

\end{document}